\documentclass[12pt]{article}
\usepackage{graphicx}
\usepackage{amsmath}
\usepackage{color}
\usepackage{cite}
\usepackage{tabularx}                                 
\usepackage{caption,subcaption}                       
\usepackage{hyperref}                                 
\usepackage[all]{hypcap}                              

\newlength{\dinwidth}
\newlength{\dinmargin}
\newcommand*{\polzraw}{{\cal P}_z}
\newcommand*{\polraw}{{\cal P}}
\newcommand*{\polz}{\ensuremath{\polzraw}}
\newcommand*{\polel}{\ensuremath{\polraw_{e^-}}}
\newcommand*{\polpo}{\ensuremath{\polraw_{e^+}}}
\newcommand*{\polzll}{\ensuremath{{\polzraw^\text{lumi}}}}
\newcommand*{\polvec}{\ensuremath{\vec{\cal P}}}                  
\newcommand*{\polvecl}{\ensuremath{\vec{\cal P}^\text{lumi,1}}}
\newcommand*{\polvecll}{\ensuremath{\vec{\cal P}^\text{lumi}}}
\newcommand*{\pol}{\ensuremath{{|\polvec|}}}                      
\newcommand*{\poll}{\ensuremath{\pol^\text{lumi,1}}}

\newcommand*{\thrraw}{\theta_r}                                   
\newcommand*{\thr}{\ensuremath{\thrraw}}
\newcommand*{\thp}{\ensuremath{\vartheta_\text{pol}}}             
\newcommand*{\thb}{\ensuremath{\vartheta_\text{bunch}}}           
\newcommand*{\xis}{\ensuremath{\xi_\text{spin}}}                  
\newcommand*{\xio}{\ensuremath{\xi_\text{orbit}}}                 
\newcommand*{\lumi}{\ensuremath{{\cal L}}}                        

\newcommand*{\br}[1] {\left( #1 \right)}
\newcommand*{\brbig}[1] {\big( #1 \big)}                          
\newcommand*{\brBig}[1] {\Big( #1 \Big)}                          
\newcommand*{\brbigg}[1]{\bigg(#1 \bigg)}                         
\newcommand*{\mean}[1] {\left\langle #1 \right\rangle}            

\newcommand*{\samnom}               {no M}
\newcommand*{\sammfive}             {M5}
\newcommand*{\sammten}              {M10}

\newcommand*{\samrdr}               {RDR}
\newcommand*{\samtdr}               {TDR}
\newcommand*{\samtdrstar}           {TDR*}
\newcommand*{\samrdrfixedEnergy}    {RDR$_0$}
\newcommand*{\samtdrfixedEnergy}    {TDR$_0$}
\newcommand*{\samtdrstarfixedEnergy}{TDR*$\hspace{-1.7mm}_0$}

\begin{document}

\newcommand {\gapprox}
   {\raisebox{-0.7ex}{$\stackrel {\textstyle>}{\sim}$}}
\newcommand {\lapprox}
   {\raisebox{-0.7ex}{$\stackrel {\textstyle<}{\sim}$}}
\def\gsim{\,\lower.25ex\hbox{$\scriptstyle\sim$}\kern-1.30ex%
\raise 0.55ex\hbox{$\scriptstyle >$}\,}
\def\lsim{\,\lower.25ex\hbox{$\scriptstyle\sim$}\kern-1.30ex%
\raise 0.55ex\hbox{$\scriptstyle <$}\,}

%
%

\begin{titlepage}
\begin{flushleft}
 {\tt DESY 14-071    \hfill    ISSN 0418-9833} \\
 {\tt May 2014}                  \\
\end{flushleft}

\vspace{1.0cm}

\begin{center}
\begin{Large}

 {\bfseries \boldmath Spin Transport and Polarimetry in the Beam Delivery System of the International Linear Collider}

\vspace{1.5cm}

M.~Beckmann$^{1,2}$, J.~List$^{1}$, A.~Vauth$^{1,2}$, and B.~Vormwald$^{1,2}$.

\end{Large}

\vspace{.3cm}
1- Deutsches Elektronen-Synchrotron DESY\\
   Notkestr. 85 \\
   22607 Hamburg, Germany
\vspace{.1cm}\\
2- University of Hamburg\\
   Institute for Experimental Physics\\
   Luruper Chaussee 149\\
   22761 Hamburg, Germany

\end{center}

\vspace{1cm}

\begin{abstract}
Polarised electron and positron beams are key ingredients to the physics programme of future linear colliders.
Due to the chiral nature of weak interactions in the Standard Model - and possibly beyond - the knowledge of
the luminosity-weighted average beam polarisation at the $e^+e^-$ interaction point is of similar importance
as the knowledge of the luminosity and has to be controlled to permille-level precision in order to fully
exploit the physics potential. The current concept to reach this challenging goal combines measurements from
Laser-Compton polarimeters before and after the interaction point with measurements at the interaction point.
A key element for this enterprise is the understanding of spin-transport effects between the polarimeters and
the interaction point as well as collision effects. We show that without collisions, the polarimeters can be
cross-calibrated to 0.1\,\%, and we discuss in detail the impact of collision effects and beam parameters on
the polarisation value relevant for the interpretation of the $e^+e^-$ collision data.

\end{abstract}

\vspace{1.0cm}

\begin{center}
\end{center}

\end{titlepage}



\section{Introduction}        \label{sec:intro}
Beam polarisation is a key ingredient to the physics programme of future linear colliders.
In particular the International Linear Collider~\cite{tdr} foresees in its baseline configuration
longitudinal polarisation of the electron and positron beams of $ \polz=\pm 80\,\%$ and $\pm 30\,\%$,
respectively. Upgrade options comprise a higher degree of positron polarisation of up to $\pm 60\,\%$.

Due to the chiral nature of weak interactions in the Standard Model, most cross-sections depend linearly on
the longitudinal beam polarisations, rendering the knowledge of the luminosity-weighted average polarisations
at the $e^+e^-$ interaction point as important as the knowledge of the luminosity itself. This does primarily
apply to electroweak precision measurements and indirect searches for new physics, e.g.\ via anomalous
couplings of the top quark~\cite{Amjad:2013tlv, powerReport}. Moreover, it also concerns direct searches for
small signals above large irreducible backgrounds, e.g.\  WIMP Dark Matter searches in the mono-photon
signature~\cite{Bartels:2012ex}. While the luminosity is expected to be measured to a few permille based on
small-angle Bhabha scattering~\cite{Bozovic-Jelisavcic:2013aca}, permille-level precision is challenging for
polarimetry. The design goal is thus to reach $\delta\polz/\polz=0.25\,\%$, while physics would profit
further if $0.1\,\%$ could be reached.

The overall concept to determine the beam polarisation is based on the interplay of several complementary
approaches: Laser-Compton polarimeters will monitor the instantaneous polarisation in some distance from the
$e^+e^-$ interaction point (IP). Their fast measurements give important feed-back to the accelerator
operators, but also track time variations on longer time scales as well as possible patterns within a
bunch train. On each beam, there will be one polarimeter $\sim 1650$\,m upstream, and a second one
$\sim 150$\,m downstream of the IP. While the upstream polarimeter measures the initial polarisation under
very clean conditions, the downstream polarimeter serves a double purpose: In absence of collisions, the
polarimeters can be cross-calibrated, if the spin transport between both locations can be predicted with
sufficient precision, i.e.\ to $0.1\,\%$ or better.
In collision mode, the depolarising effects of the beam-beam interaction and the luminosity-weighted average
polarisation at the IP could be monitored. This enterprise, however, depends crucially on a sufficiently
precise understanding of all effects of the beam-beam interaction and the spin transport, which is the main
objective of this publication.

The luminosity-weighted average polarisation at the IP can be directly accessed from $e^+e^-$ collision data
themselves. Several approaches have been studied for this in the
past~\cite{Moenig, thesisIvan, EPSprocedings}, comprising schemes operating on the
measurements of total cross-sections for various polarisation configurations, as well as on single- and
double-differential distributions of $W^+W^-$ production. All these approaches will finally yield a very
important long-term scale calibration of the luminosity-weighted average polarisation when appropriately
compared with the polarimeter measurements. However, they all will take years of data taking before reaching
permille-level precision. Furthermore, they assume that the helicity reversal is exact, i.e.\ that the
absolute value of the polarisation is the same for all helicity configurations. Any deviation from this
assumption can only be corrected for based on the polarimeter measurements and their propagation to the IP.
It has been shown that the need for such a correction could limit the precision of the collision data methods at the precision of the
polarimeters~\cite{thesisIvan}. The difference in absolute polarisation values between data sets, however, can
be minimized if the helicity of both beams can be reversed quickly and independently, e.g.\ on a
pulse-by-pulse basis~\cite{Riemann:2009wy}. For the electron beam, this is readily provided by switching the
helicity of the source laser. For the positron beam, currently a scheme based on switching between two spin
rotator beam lines as proposed in~\cite{Riemann:2012zza} is foreseen~\cite{tdr}.

This paper is organised as follows: Sections~\ref{sec:bds} and~\ref{sec:sim} introduce the beam delivery
system of the ILC  and the spin-tracking formalism, respectively. The design and the capabilities of the
Laser-Compton polarimeters will be summarised in section~\ref{sec:polarimetry}. In section~\ref{sec:crosscal}
we will study the achievable precision for cross-calibration of the polarimeters in the absence of
collisions, while section~\ref{sec:lumipol} discusses the extraction of the luminosity-weighted average
polarisation in the presence of collisions. We conclude in section~\ref{sec:conclusions} with the prospects
for reaching the goal of a few permille precision and give an outlook on studies required in the future.

\section{Accelerator Environment}       \label{sec:bds}
In this section the parts of the ILC most relevant for the determination of the luminosity-weighted average
polarisation at the $e^+e^-$ interaction point will be introduced. This includes the beam delivery systems
and the extraction lines as well as the different beam parameter sets used in the simulation study.
Since most of the aspects studied in this paper are identical between electrons and positrons, we will use
the term ``electron'' for both beams, unless specific differences need to be pointed out.

\subsection{The ILC Beam Delivery System and Extraction Line} \label{subsec:ilcbds}
\begin{figure}[hbt]
   \centerline{
      \includegraphics*[width=0.75\textwidth]{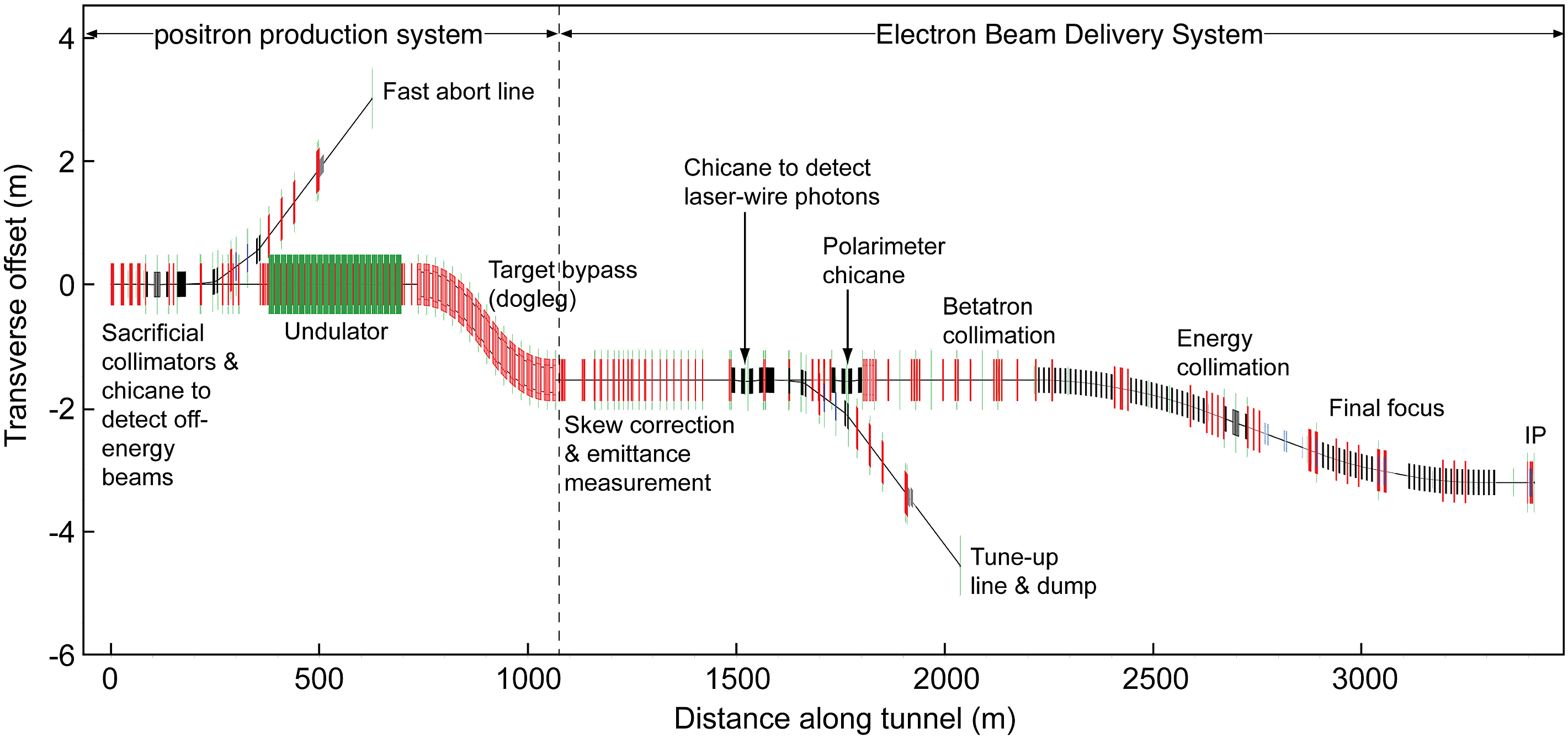}
   }
   \caption[]{Layout of the electron beam delivery system (BDS). The positron BDS is a mirror image.
              From figure~2.12 in~\cite{tdr32}.}
   \label{fig:bds-SB2009}
\end{figure}

The beam delivery system (BDS) is an about $2$\,km long set of beamlines which serves the final preparation of the
fully accelerated beams for collisions. It hosts beam diagnostics, skew correction, betatron and energy
collimation as well as the final focus system, and has been carefully designed to minimise the emittance
growth. This applies in particular to the vertical direction, since flat beams are essential to reconcile
high luminosity with minimal beamstrahlung~\cite{thesisschulte}. Therefore, all bends in the BDS are in the horizontal plane.
Figure~\ref{fig:bds-SB2009} shows a sketch of the BDS for the electron beam up to the $e^+e^-$ IP. The
upstream polarimeter is located directly behind the branch-off to the tune-up dump. The positron BDS is
a mirror image, apart from the positron production system (left part of figure~\ref{fig:bds-SB2009}).

\begin{figure}[hbt]
   \centerline{
   \includegraphics*[width=0.75\textwidth]{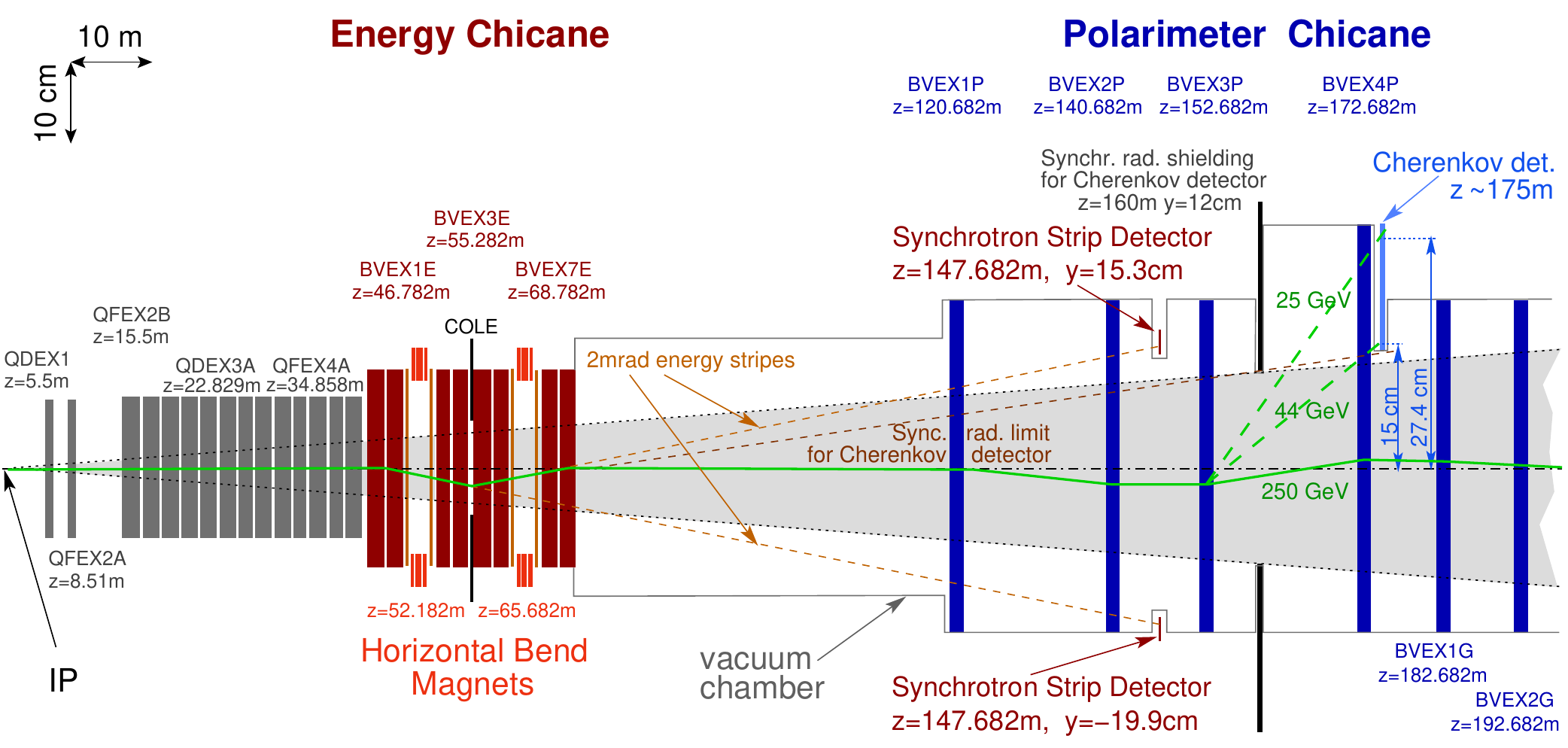}
   }
   \caption[]{Schematic view of the extraction line up to the downstream polarimeter.
   From~\cite{boogert}.}
   \label{fig:downstreamPolChicane}
\end{figure}

Figure~\ref{fig:downstreamPolChicane} shows the extraction line, which guides the spent beams to the dumps
and provides post-collision diagnostics. In particular, the downstream polarimeter is located at a secondary
focus point of the optics, in order to provide the best possible beam conditions at the Compton IP of the
downstream polarimeter.

\subsection{Beam Parameters and Interaction Region}       \label{subsec:beampars}

\begin{table}[hbt]
   \centering
   \begin{tabular}{lllcc}
      Parameter                       & \multicolumn{2}{l}{Symbol}              & RDR        & TDR         \\
      \hline
      Bunches per train               &                       &                 & $2\,625$   & $1\,312$    \\
      Train frequency                 &                       & [Hz]            & $5$        & $5$         \\
      Horizontal bunch size           & $\sigma_{xe}$         & [nm]            & $639$      & $474$       \\
      Vertical   bunch size           & $\sigma_{ye}$         & [nm]            & $5.7$      & $5.9$       \\
      Horizontal angular spread       & $\theta_x$            & [$\mu$rad]      & $32$       & $43$        \\
      Vertical   angular spread       & $\theta_y$            & [$\mu$rad]      & $14$       & $12$        \\
      Horizontal norm. emittance      & $\gamma\varepsilon_x$ & [$\mu$m]        & $10$       & $10$        \\
      Vertical   norm. emittance      & $\gamma\varepsilon_y$ & [nm]            & $40$       & $35$        \\
      Horizontal disruption parameter & $D_x$                 &                 & $0.17$     & $0.3$       \\
      Vertical   disruption parameter & $D_y$                 &                 & $19.4$     & $24.6$      \\
      Beam energy spread ($e^-, e^+$) & $\sigma_E/E$          & [$10^{-3}$]     & $1.4, 1.0$ & $1.24, 0.7$ \\
      $e^+e^-$ luminosity             & $\cal L$  & [$10^{34}$\,cm$^{-2}$\,s$^{-1}$] & $2$   & $1.8$       \\
   \end{tabular}
   \caption[]{Selected beam parameters at the IP for $E_{cm}=500\,$GeV according to RDR~\cite{rdr3} (nominal parameter
              set) and TDR~\cite{tdr32} (baseline parameters).}
   \label{tab:beamParametersDesign}
\end{table}

Table~\ref{tab:beamParametersDesign} lists the design beam parameters at the IP according to the Reference
Design Report (RDR) \cite{rdr} and the more recent Technical Design Report (TDR) \cite{tdr}. Most of the
studies presented in this paper have been performed with the RDR parameters, since they match the available lattice\footnote{The \textit{lattice} describes the layout of a beamline, e.g.\ positions and strengths of
the magnets.}.
For the current TDR parameter set, the number of bunches per train is reduced with respect to the RDR.
In order to restore the luminosity, the beams are focussed more strongly at the IP. This results in more intense collisions, which might also affect the polarsation. Therefore,
the collision effects and the spin transport to the downstream polarimeter have been studied for both
parameter sets. The electron and positron beam parameters are identical apart from the beam energy spread
$\sigma_E/E$, which is slightly increased for the electrons during their passage through the undulator of
the positron source.

The ILC design foresees a horizontal crossing angle of $14\,$mrad between the $e^-$ and the $e^+$ beamlines at the IP.
In order to maximise the luminosity, the bunches are correspondingly rotated by $7\,$mrad using
crab cavities~\cite{Adolphsen:2007zza}.

The collider experiments at the IP contain two types of magnets which also influence the beams. These are the
main solenoid for the tracking system and the anti-DID (detector-integrated dipole) to guide
electron-positron pairs produced by beamstrahlung photons into the outgoing beam pipe~\cite{Seryi:2006ja}. Due to the crossing
angle, the detector magnets at the IP are rotated by $7\,$mrad with respect to each of the beamlines.
These magnets are not yet present in the official lattice files, but have been included in our simulations
by hand.

\subsection{Magnet Misalignments and Orbit Correction}       \label{subsec:misalignments}
Misalignments of the magnets in an accelerator have various effects, among others beam jitter (deviations of
the beam orbit from the design orbit) or a detuned focussing. There are static misalignments due to the
limited precision to which the beamline elements can be adjusted at their designated positions, and
time-dependent misalignments from various sources, e.g.\ from seismic noise, traffic, or cooling water pumps.

To minimise the beam jitter and prevent a beam loss in the worst case, the ILC will be equipped with several
feedback orbit correction systems. These systems consist of dipole magnets to perform the orbit correction
and of beam position monitors, which measure the current beam position from which the required field
strengths for the correction magnets are recalculated. They operate on a timescale of a tenth of a second,
i.e.\ the measurements from one bunch train are used to correct the orbit for the next bunch train. At the IP,
a higher precision is required to bring the beams with vertical sizes of few nanometers to collision with the
envisaged luminosity. Therefore, an additional fast-feedback system is foreseen at the IP, which operates
on a timescale of nanoseconds, i.e.\ bunch-to-bunch. The interplay of ground motion and the fast-feedback
system and their effects on luminosity and polarisation have been studied
previously~\cite{RestaLopez:2008zza, Bailey:2011ey}. It has been shown in particular that without correction,
ground-motion-induced misalignments and the resulting depolarisation become significant only at time scales
of a day. Thus, the feedback required for maintaining luminosity is considered sufficient to also maintain
polarisation, and we will not study time-dependent misalignments in this publication.

\section{Spin Transport and Collision Effects}       \label{sec:sim}
In this section, the basic effects on the polarisation which occur in the beam delivery system, the
interaction region, and the extraction line as well as their implementation in the simulation are introduced.

\subsection{Spin Transport}       \label{subsec:spinTransport}
The spins of particles in an accelerator are subject to spin precession in electromagnetic fields and to
spin-flips from the emission of photons. For the bunch propagation through the ILC beam delivery system,
it turns out that the effects of spin-flips can be neglected (cf.\ section~\ref{sec:crosscal}). The energy
loss due to synchrotron radiation, however, affects the particle trajectories and the spin precession
downstream.

The change of a particle spin vector $\vec{S}$ with time under the influence of electromagnetic fields is
described by the Thomas-Bargmann-Michel-Telegdi (T-BMT) equation~\cite{Thomas:1926dy, Bargmann:1959gz}.
Since neither the beam delivery system nor the extraction line contains components with sizable electric
fields, the full T-BMT equation simplifies to
\begin{equation}
\label{eqn:TBMT}
   \dfrac{d}{dt} \vec{S} = \vec{\Omega}_B\br{\vec{B}, \vec{r},\vec{p},t} \times \vec{S}
                         = -\dfrac{q}{m\gamma}\br{ \br{1+a\gamma}\vec{B}
                           - \dfrac{a\,\vec{p}\cdot\vec{B}}{(\gamma+1)\, m^2 c^2}\,\vec{p} } \times \vec{S}.
\end{equation}

Here, $\vec{B}\br{\vec{r},t}$ denotes the magnetic field, $\gamma$ the relativistic Lorentz factor,
$c$ the vacuum speed of light and $a \equiv (g-2)/2$ the anomaly of the gyro-magnetic moment, with
$a \approx 0.001159652$ for electrons.

The expression for $\vec{\Omega}_B$ in equation~\ref{eqn:TBMT} can be decomposed in two parts for the field
components, $\vec{B}_\parallel$ parallel to $\vec{p}$ and $\vec{B}_\perp$ perpendicular to it:
\begin{equation}
   \vec{\Omega}_B\br{\vec{B}, \vec{r},\vec{p},t} = -\dfrac{q}{m\gamma}\brbigg{\br{1+a\gamma}\vec{B}_\perp\;\; +\;\;\br{1+a}\vec{B}_\parallel}
   \label{eq:tbmtOmegaBdecomposed}
\end{equation}

In presence of only perpendicular magnetic fields, the momentum $\vec{p}$ and the spin $\vec{S}$ behave very
similarly:
\begin{align}
   \dfrac{d}{dt} \vec{p} =& -\dfrac{q}{m\gamma}\br{\phantom{\br{1+a\gamma}}\vec{B}_\perp} \times \vec{p}\\
   \dfrac{d}{dt} \vec{S} =& -\dfrac{q}{m\gamma}\br{         \br{1+a\gamma} \vec{B}_\perp} \times \vec{S}
\end{align}

Thus, the spin vector precesses in a perpendicular magnetic field about $\vec{B}$ by the angle
\begin{equation}
   \xis = \br{1+a\gamma}\,\xio,
   \label{eq:tbmtTheta}
\end{equation}
where \xio\ is the deflection angle of the particle~\cite{hoffstaetter,thesisjeff}.
For an electron beam with an energy of $250\,$GeV, the amplification factor is
$\br{1+a\gamma}\approx 568$; for an energy of $500\,$GeV, it rises to $\approx 1136$.

\begin{figure}[ht]
   \begin{center}
      \includegraphics[width=0.5\columnwidth]{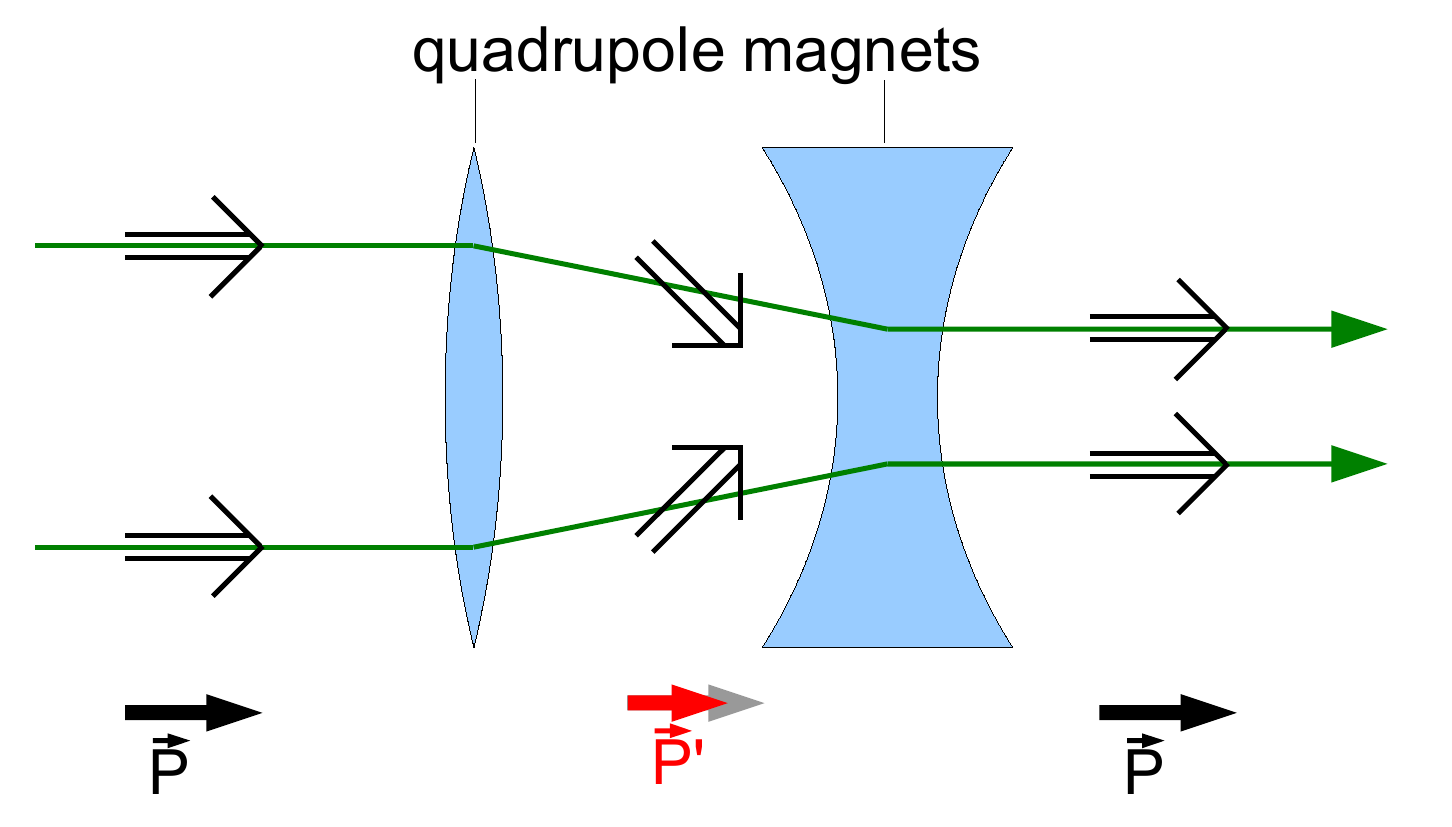}
   \end{center}
   \caption{Spin fan-out at the example of two hypothetical particles. This sketch only illustrates the
            possible effects of quadrupole magnets on the beam polarisation. It does not describe realistic
            focussing, the two shown quadrupole magnets do not have the same strength, and betatron
            oscillations in two dimensions are not indicated. \cite{thesisMoritz}}
   \label{fig:spinFanOut}
\end{figure}

For a particle beam with a spatial extension, an angular divergence and an energy spread in an inhomogeneous
magnetic field, the spin precessions for different particles can vary significantly.
One possible pattern emerging from non-uniform spin precession is referred to as ``spin fan-out'' in the following
and occurs in inhomogeneous magnetic fields or in the presence of a beam energy spread.
Figure~\ref{fig:spinFanOut} illustrates spin fan-out and its reversability using the example of a hypothetical two-particle bunch traversing
quadrupole magnets: Both particles are deflected into different directions in the first quadrupole and the
spin vectors precess correspondingly according to equation~\ref{eq:tbmtTheta}. Thus, the longitudinal
polarisation decreases, but the transverse polarisation remains zero since the transverse components of the
two spin vectors cancel each other. Consequently, the magnitude of the polarisation decreases:
$|\polvec'|<|\polvec|$. This can in principle be reversed by a second quadrupole which rotates the spin
vectors back to the original orientation.

In a beamline consisting of F0D0 cells\footnote{Standard set of focussing (F) and defocussing (D) quadrupoles
interleaved with drift spaces (0).}, however, these spin precessions occur alternatingly in the horizontal and
the vertical plane. Rotations in two dimensions do generally not commute, but this fact can be neglected if
the rotations are sufficiently small. In that case, the spin fan-out can be set into relation to the angular
divergence \thr\ (i.e.\ to the fan-out of the momentum vectors) of the bunch by the following function,
assuming a longitudinally polarised beam, for which the maximum polarisation $\pol_\text{max}$ is obtained
for $\thr=0$ (as in figure~\ref{fig:spinFanOut}):
\begin{equation}
   f(\thr)=\pol_\text{max}\cdot\cos\brBig{\brbig{1+a\mean{\gamma}}\cdot\thr},
   \label{eq:fThr}
\end{equation}
where $\mean{\gamma}$ denotes the average relativistic Lorentz factor.

In presence of a beam energy spread, a similar behavior occurs (also in homogeneous magnetic fields),
since the ``amplification factor'' $\br{1+a\gamma}$ is energy-dependent.
This fan-out is in principle reversible as well.

So it might seem that polarisation can be generated by spin fan-out, but that is actually not the case.
Spin fan-out allows only to restore an existing ordering in the spin orientation which is preserved in a
correlation between the spin orientation and the particle energy or a particle coordinate. It is not
possible to restore polarisation which got lost in stochastic processes like radiative depolarisation,
which will be introduced in the following section.

The description of the spin fan-out by equation~\ref{eq:fThr}, however, relies on sufficiently small rotations.
Verifying the validity of this assumption for the ILC BDS is one of the objectives of a full spin tracking study.

\subsection{Beam-Beam Collisions and Luminosity-Weighted Average Polarisation}
\label{subsec:beambeamCollisions}
In the beam-beam collisions at the IP, the colliding bunches distort each other by their electromagnetic
fields, while only a few particles undergo hard interactions. The polarisation of these particles is the
luminosity-weighted polarisation for a single collision of an electron bunch with a positron bunch, which
will be denoted here by the symbol \polvecl. It is luminosity-weighted with respect to the local distribution
of the luminosity during a collision.

The decisive quantity for interpreting collision data is the luminosity-weighted polarsation averaged over
time \polvecll\ for each of the beams, defined as
\begin{equation}
   \polvecll = \dfrac{\int \lumi(t) \polvecl(t)\,dt}{\int \lumi(t)\,dt},
   \label{eqn:pollumi}
\end{equation}
and in particular its longitudinal component $\polzll$. Depending on the chiral structure of the studied
observable, it might also be advantageous to consider the time average of one of the effective
polarisations~\cite{Moenig} listed in table~\ref{tab:effpol} instead, which can all be defined in analogy to
equation~\ref{eqn:pollumi}.
\begin{table}[hbt]
   \centering
   \begin{tabular}{c|l|l}
   effective polarisation & observable & type of interaction\\
   \hline
   $\frac{\polel + \polpo}{1 + \polel \polpo}$ & $A_{LR}$ & $s$-channel vector exchange \\
   $\polel \polpo$ & cross-section & $s$-channel vector exchange \\
   $\polel + \polpo - \polel \polpo$ & cross-section & $t$-channel $W$ or $\nu_e$ production \\
   \hline
   \end{tabular}
   \caption[]{Effective polarisations minimising the impact of polarisation uncertainties on various observables.}
   \label{tab:effpol}
\end{table}
The gain in error reduction when using the appropriate effective polarisations depends on the
degree of correlation between the measurements of electron and positron polarisation. In this
respect, it is reasonable to assume that the instrumental uncertainties of the polarimeters
as well as the influence of misalignments in the BDS are uncorrelated. Collision effects on
the other hand have to be assumed to be highly correlated, since the intensity of the collision is reciprocal.

In head-on collisions, the already focussed bunches attract each other even further due to their mutual electrical fields (pinch effect~\cite{thesisschulte}).
The T-BMT precession due to this mutual focussing of the bunches leads to a spin fan-out like in a quadrupole
magnet. For flat, longitudinally polarised beams ($\sigma_{xe}\gg\sigma_{ye}$, $\polz = \pol$) and a small
horizontal disruption parameter ($D_x\ll 1$), the polarisations before ($\pol^\text{bef}$) and after ($\pol^\text{aft}$) the collision as well as the luminosity-weighted polarisation \poll\ are related as follows
(equation~16 in \cite{yokokaChen}):
\begin{equation}
   \pol^\text{bef}-\poll = 0.273 \br{\pol^\text{bef}-\pol^\text{aft}}
   \label{eq:yokoyaChen16}
\end{equation}

If the angular divergence of the bunches before the collision is negligible in terms of spin fan-out, the
spin fan-out during the collision can be related to the angular divergence $\thrraw^\text{aft}$ after the
collision (equation~31 in \cite{yokokaChen}, see also equation~\ref{eq:fThr}):
\begin{equation}
   \pol^\text{bef}-\pol^\text{aft} \approx \dfrac{1}{2} \pol^\text{bef} \cdot\br{1+a\gamma}^2 \cdot(\thrraw^\text{aft})^2
   \label{eq:yokoyaChen31}
\end{equation}

Merging these two equations, one obtains:
\begin{equation}
   \pol^\text{bef}-\poll \approx \dfrac{1}{2} \pol^\text{bef} \cdot\br{1+a\gamma}^2 \cdot\br{\dfrac{\thrraw^\text{aft}}{2}}^2
   \label{eq:mikesPaper}
\end{equation}
As explained in \cite{mikesPaper} and \cite{moenig}, one can interpret this as about half of the T-BMT precession occurring
before the hard interaction. A comparison to equation~\ref{eq:yokoyaChen31} implies that one can reproduce
the luminosity-weighted polarisation at a point behind the IP where the angular divergence has to be reduced
by a factor $1/2$ with respect to the divergence at the IP after the collision.

During the mutual distortion of the bunches, the deflected particles also emit beamstrahlung photons, which
can cause a flip of the electron spin by the Sokolov-Ternov effect~\cite{Sokolov:1963zn}. Unlike for the
synchrotron radiation in the BDS, the radiative depolarisation in the collision could reach a non-negligible level. Therefore, dedicated simulations of the bunch-bunch interaction are needed in
order to verify whether equation~\ref{eq:mikesPaper} holds for realistic ILC beam parameters.
A detailed description of the beamstrahlung-related effects on the polarisation can be found in section~5.9 of \cite{cainManual}.

\subsection{Simulation Framework {\tt STALC}}      \label{subsec:simulationFramework}
In order to study the spin transport in the beam delivery system of a linear collider, the simulation framework
{\tt STALC} (Spin Transport at Linear Colliders) has been developed \cite{thesisMoritz}. As sketched in
figure~\ref{fig:simulationFramework}, it interfaces dedicated programs to simulate the beam transport
including the polarisation through the BDS, the collisions at the IP and the measurements of the
polarimeters, respectively:

\begin{figure}[ht]
   \includegraphics[width=\columnwidth]{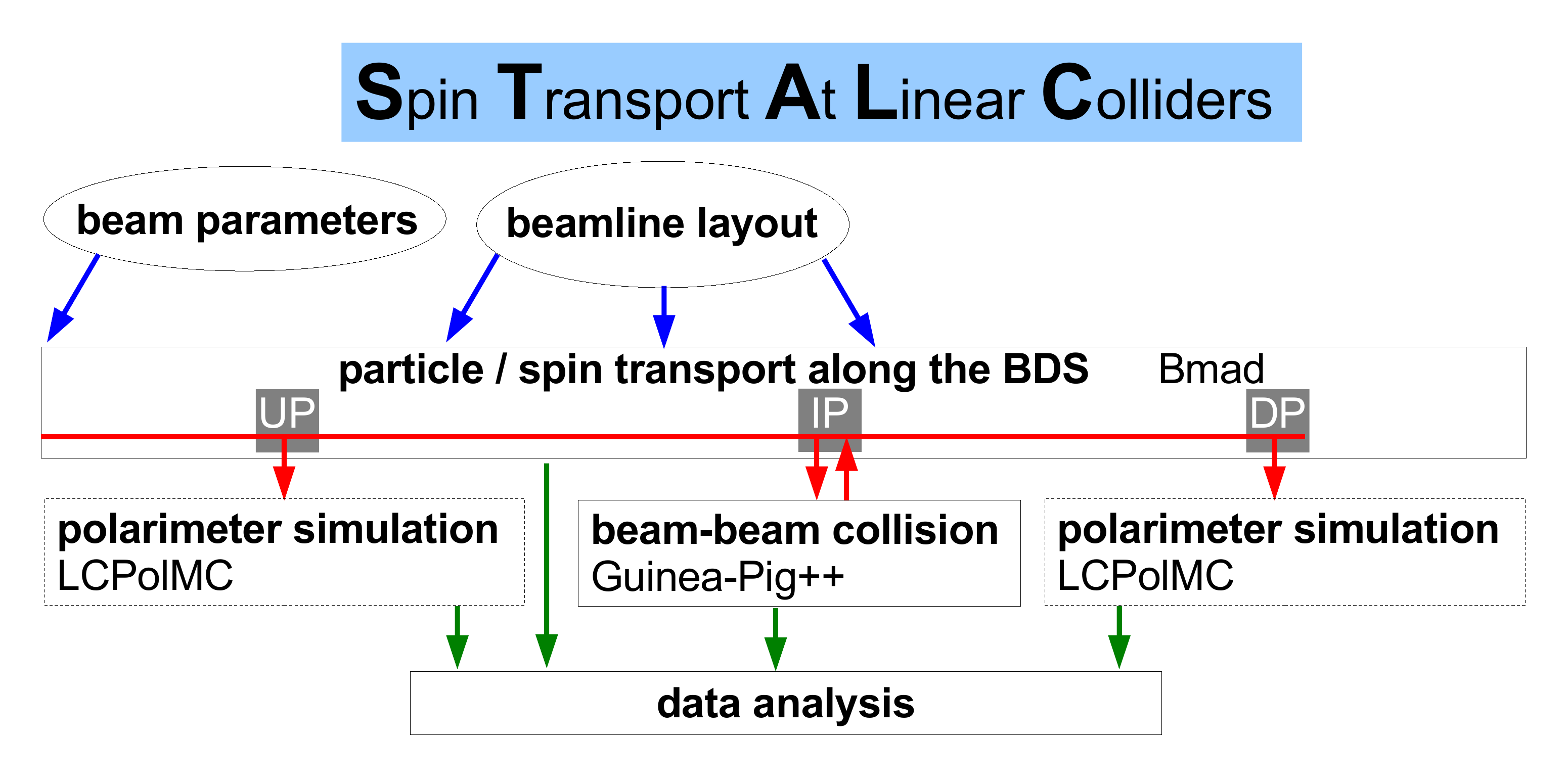}
   \caption{Program flow in {\tt STALC}. UP/DP denotes the up-/downstream polarimeter.}
   \label{fig:simulationFramework}
\end{figure}

{\tt STALC} generates electron/positron bunches and tracks them through a given lattice via
{\tt Bmad}~\cite{bmad,bmadmanual}, a subroutine library for particle simulations in high-energy accelerators,
which can take also the polarisation into account. At the polarimeters, the particle information can be
passed to {\tt LCPolMC}, which simulates the polarisation measurement in a Compton polarimeter from the
Compton scattering process to the detector response. At the IP, the particle information from one electron
bunch and one positron bunch is passed to {\tt Guinea-Pig++}~\cite{guineapig++} to simulate the beam-beam
collisions at the IP. {\tt Guinea-Pig++} is an extension of {\tt Guinea-Pig}~\cite{guineapig}, which takes
into account the polarisation and simulates T-BMT precession and Sokolov-Ternov effects. The spent bunches
after the collision are passed back to the accelerator simulation and tracked to the downstream polarimeter.
The final analysis is based on {\tt ROOT}~\cite{root}.

{\tt STALC} is not inherently limited to the ILC, but can be run on any lattice and beam parameter set.

\section{Compton Polarimetry at the ILC}       \label{sec:polarimetry}
In this section, we will summarise the working principle and the capabilities of the Laser-Compton polarimeters
at the ILC. A more detailed description of both polarimeters can be found in~\cite{boogert}.

The single differential cross section $d \sigma / dE$ for Compton scattering~\cite{fano} contains a term
proportional to the product of the longitudinal beam polarisation and the circular laser polarisation, which
can be isolated by measuring the count-rate asymmetry with respect to the laser helicity.

In order to quickly gather statistical precision, both polarimeters operate in a multi-event
mode with $\mathcal{O}(10^3)$ Compton interactions per bunch.

\begin{table}[hbt]
   \centering
   \begin{tabular}{lccc}
                              & $e^+/e^-$ beam      & \multicolumn{2}{c}{Laser beam} \\
                              &                     & Upstream         & Downstream  \\
      \hline
      Energy                  & $45.6$ - $500$\,GeV &  $2.33$\,eV      & $2.33$\,eV  \\
      Bunch charge/energy     & $2\cdot10^{10}\,e$  &  $35\,\mu$J      & $100$\,mJ   \\
      Bunches per train       & $1312$ - $2625$     &  $1312$ - $2625$ & $1$         \\
      Bunch length $\sigma_t$ & $1.3$\,ps           &  $10$\,ps        & $2$\,ns     \\
      Average power           &                     &  $0.2$-$0.5$\,W  & $0.5$\,W    \\
   \end{tabular}
   \caption[]{Selected parameters of the electron and laser beams at the up- and downstream polarimeter.}
   \label{tab:polbeampars}
\end{table}

Table~\ref{tab:polbeampars} summarises specifications of the electron and laser beams relevant for the
luminosity calculation. A striking difference between the up- and downstream polarimeter lasers is the energy
per bunch, which differs by more than three orders of magnitude.
At the location of the downstream polarimeter, significant amounts of background are expected, at the level
of ${\cal{O}}(10^3)$ photons and ${\cal{O}}(10^2)$ charged particles per bunch
crossing~\cite{Ken:EPWS08proceedings}. In order to maintain a suitable signal-to-background ratio in
presence of this background, a significantly higher laser power per shot is needed, at the price of a much
lower repetition rate and longer pulse durations. This allows to shoot at one electron bunch per train, or
a few if several laser are employed. In the nearly background-free conditions of the upstream polarimeter,
it is possible to cover every bunch in a train, e.g.\ by employing similar lasers as for the electron
source~\cite{teslaPolarimeter}.

\begin{table}[hbt]
   \centering
   \begin{tabular}{lccccccc}
                                          &           & \multicolumn{2}{c}{UP} & \multicolumn{2}{c}{DP} &  DP with collisions\\
                                          &           & $e^+$   & $e^-$        & $e^+$ & $e^-$          & $e^+/e^-$          \\
      \hline
      Horizontal bunch size $\sigma_{xe}$ &[$\mu$m]   & $24$    & $32$         & $7$   & $15$           & $\sim 3000$        \\
      Vertical   bunch size $\sigma_{ye}$ &[$\mu$m]   &  $3$    &  $3$         & $33$  & $39$           & $\sim 1200$        \\
      Beam energy spread $\sigma_E/E$     &[$10^{-3}$]& $0.7$   & $1.2$        & $0.7$ & $1.3$          & $\sim 44$          \\
   \end{tabular}
   \caption[]{Size and energy spread of electron and positron beams at the up- and downstream polarimeter locations obtained from {\tt STALC}.}
   \label{tab:polbeamsizes}
\end{table}

The relevant electron beam sizes at the polarimeter location as obtained from {\tt STALC} are shown in
table~\ref{tab:polbeamsizes}.
It should be noted that after collisions, the beams are highly disrupted and have large non-Gaussian tails,
as will be discussed in more detail in section~\ref{subsec:measurablePolarisation}.
Thus, the Gaussian beam sizes listed here should be considered as indicative only.

\begin{table}[hbt]
   \centering
   \begin{tabular}{lccc}
                                                          &                                & Upstream      & Downstream      \\
      \hline
      Beam crossing angle  $\alpha$                       & [mrad]                         & $10$          & $15.5$          \\
      Crossing plane                                      &                                & horizontal    & vertical        \\
      Laser beam size $\sigma_{x\gamma}=\sigma_{y\gamma}$ & [$\mu$m]                       & $50$          & $100$           \\
      Luminosity / bunch                                  & [$10^{4}$\,b$^{-1}$]           & $1.0$         & $18$            \\
      Luminosity                                          & [$10^{29}$\,cm$^{-2}$s$^{-1}$] & $630$         & $9.1$           \\
      $\delta \polz / \polz$ (stat)                       &                                & $<1\,\% /$\,s & $<1\,\% /$\,min \\
      $\delta \polz / \polz$ (sys)                        &                                & $0.25\,\%$    & $0.25\,\%$      \\
   \end{tabular}
   \caption[]{Selected parameters of the up- and downstream polarimeters. For the downstream polarimeter, the luminosities 
   are based on the electron beam parameters in absence of 
   collisions at the $e^+e^-$ IP.}
   \label{tab:polpars}
\end{table}

Table~\ref{tab:polpars} finally shows the actual luminosities per bunch and per second, calculated for laser
crossing under a small angle $\alpha$ in the vertical plane according to
\begin{equation}
{\cal{L}} = \frac{N_e N_{\gamma}}{2\pi \sqrt{(\sigma_{xe}^2 + \sigma_{x\gamma}^2)}\sqrt{(\sigma_{ye}^2 + \sigma_{y\gamma}^2)+(\sigma_{ze}^2 + \sigma_{z\gamma}^2)(\alpha/2)^2}}.
\end{equation}

As desired, the luminosity per bunch turns out to be a factor~$20$ larger at the downstream polarimeter than
at the upstream polarimeter, at the price of a much lower instantaneous luminosity. This in turn leads to a
longer time required to reach the same statistical precision. Nevertheless, both measurements
are expected to be systematically limited after a very short time.

Since the electron beam is much more energetic than the (optical) laser photons, all scattered particles are
forward directed within a narrow cone of $10-20\,\mu$rad. For the relevant photon and electron energies, the
total Compton cross section amounts to $154$ ($118$)\,mb for anti-parallel (parallel)  laser and electron
polarisation. For the luminositiy of the upstream
(downstream) polarimeter, this results in ${\cal O}(10^3)$ (${\cal O}(10^4)$)  Compton scatterings per bunch.
The Compton scattered electrons are momentum-analysed by a set of dipole magnets, and their flux is measured
as a function of position. The count-rate asymmetry expected for a fully polarised beam is called analysing
power. The magnets are arranged as part of a chicane such that the undisturbed beam resumes its original
trajectory after passing the whole polarimeter section, while the Compton-scattered electrons should be
kicked out sufficiently far from the main beam axis to allow detection.

In case of the upstream polarimeter, the chicane consists of four symmetric sets of dipole magnets, with
the Compton interaction point in the middle and the detector behind the forth set of dipoles. Such a
configuration offers the additional advantage of decoupling the detector acceptance from the initial beam
energy, since the Compton spectrum is projected onto the same area in the detector plane for all beam
energies. Instead the dispersion in the middle of the chicane and thus the position of the Compton IP
changes according to beam energy.

In case of the downstream polarimeter, the chicane is based on six dipoles, again with the detector behind
the forth dipole. In this case, the third and forth dipole are operated at a larger magnetic field in order
to kick the Compton signal sufficiently far out of the synchrotron radiation fan from upstream magnets.
The fifth and sixth dipole then bring the main beam back to its original trajectory. Detailed specifications
for both chicanes can be found in~\cite{boogert}.

The lattice between the IP and the downstream polarimeter has been designed such that the angular divergence
of the beam at the downstream polarimeter is by a factor two smaller than at the end of the collision\footnote{
I.e.\ that the relevant element of the transfer matrix between $e^+e^-$ IP and the downstream polarimeter
Compton IP is $|R_{22}|=0.5$.}. Thus, the downstream polarimeter measurement is expected to relate directly
to the luminosity-weighted average polarisation during collisions as explained in
section~\ref{subsec:beambeamCollisions}. Whether this picture holds e.g.\ in presence of beamstrahlung will
be investigated in section~\ref{sec:lumipol}.

Building on the successful experience at SLD~\cite{Abe:2000dq}, the baseline foresees an array of $20$
gas-Cherenkov detectors read-out by photomultiplier tubes~\cite{Bartels:2010eb}. The advantages of this
technology comprise radiation hardness, robustness against low-energetic backgrounds due to a Cherenkov
threshold in the MeV-regime and a low number of channels to be read-out.

The limiting systematic uncertainties of the SLD polarimeter were the knowledge of the analysing power and
the photo detector non-linearities. Both aspects have been studied for the ILC polarimeters in a recent R\&D effort: The dominating
contribution to the analysing power, i.e.\ the alignment of the detector with respect to the beam, has been studied
in a testbeam campaign with a prototype~\cite{Bartels:2010eb}, and a calibration system for monitoring and
correcting non-linearities to a sufficient level has been developed~\cite{Vormwald:thesis}. Based on this
experience, the uncertainty budget listed in table~\ref{tab:polsys} seems achievable. Details on the online
monitoring of the laser polarisation can be found in~\cite{Ken:EPWS08proceedings}. The total systematic
uncertainty on the polarimeter measurements amounts to $\delta \polz/\polz = 0.25\,\%$.

\begin{table}[hbt]
   \centering
   \begin{tabular}{lc}
      Source of uncertainty            & $\delta \polz / \polz$ \\
      \hline
      Detector analysing power         & $0.15 - 0.2\,\%$       \\
      Detector linearity               & $0.1\,\%$              \\
      Laser polarisation               & $0.1\,\%$              \\
      Electronic noise and beam jitter & $0.05\,\%$             \\
      \hline
      Total                            & $0.25\,\%$             \\
   \end{tabular}
   \caption[]{Uncertainty budget for the Compton polarimeters.}
   \label{tab:polsys}
\end{table}

All the contributions in table~\ref{tab:polsys} are expected to be uncorrelated between all four
polarimeters. Thus, additional precision could be gained from a cross calibration of the polarimeters in
absence of collisions, if the spin transport from one location to the other is sufficiently well understood.
In presence of collisions, both polarimeters fulfil very complementary tasks: While the downstream
polarimeter in principle gives access to the depolarisation in collision, the upstream polarimeter provides
a clean measurement of the initial polarisation and resolves possible time dependent patterns, e.g.\ inside
a bunch train.

\section{Cross calibration of the Polarimeters}       \label{sec:crosscal}
In absence of collisions at the $e^+e^-$ IP, the polarisation measurement at the upstream polarimeter can be
propagated by spin tracking along the BDS and predict the expected polarisation
at the downstream polarimeter. Provided that the spin propagation effects are well under control,
the two polarimeters can thus be cross calibrated. However, it is crucial to understand how deviations
due to instrumental effects of the polarimeters can be disentangled from effects caused by
the spin transport. Therefore, it is important to understand the individual impact of the various 
effects which could influence the spin transport between the polarimeters. Both the
cross calibration of the polarimeters and the understanding of the spin transport are crucial
to finally assess the collision effects and to extract the luminosity-weighted polarisation.

We studied the spin transport with {\tt STALC} based on the SB2009\_Nov10 lattice \cite{SB2009lattice}, which
differs from the RDR lattice with respect to the new location of the upstream polarimeter\footnote{Due to its
separation from the emittance measurement following earlier recommendations~\cite{Kafer:2008ku}.}. For the
positrons, the tracking starts at the beginning of the BDS (dashed line in figure~\ref{fig:bds-SB2009}).
On the electron side, the simulation starts at the beginning of the positron production system,
thus including the ``dogleg''
to bypass the positron source target area. In both cases, the beams are initialised according to the twiss parameters of the lattice and the TDR
beam energy spreads. However, the final focus system of the SB2009\_Nov10 lattice was not
yet adapted to the TDR design and thus produces the RDR beta functions at the $e^+e^-$ IP.
In absence of collisions, no significant impact on the spin transport is expected due to this issue.

To avoid numerical problems, some magnets have been sliced into smaller units as described in chapter~7 of
\cite{thesisMoritz}. In each studied configuration, 1000 individual bunches have been tracked, each
bunch with 40\,000 macroparticles\footnote{Thus, for a design bunch population of $N_{e}= 2\cdot 10^{10}$,
one macroparticle represents $5\cdot10^5$ electrons.}. To each macroparticle, a polarisation vector with a
length of $\pol_\text{max}=0.8$ is assigned.
At the beginning of the tracking, the polarisation vectors are assumed to be oriented such that they are
parallel to the orbit if the macroparticle trajectory is parallel to the orbit as well, and rotated
analogously to T-BMT precession otherwise. In this case, the spin fan-out is expected to be described by
equation~\ref{eq:fThr}. This spin configuration has already implicitly been assumed in earlier studies
\cite{yokokaChen, spinConfigExample}. The effects of a slightly different spin configuration will be
examined in section~\ref{sec:lumipol}.

\begin{figure}[ht]
   \includegraphics[width=\columnwidth]{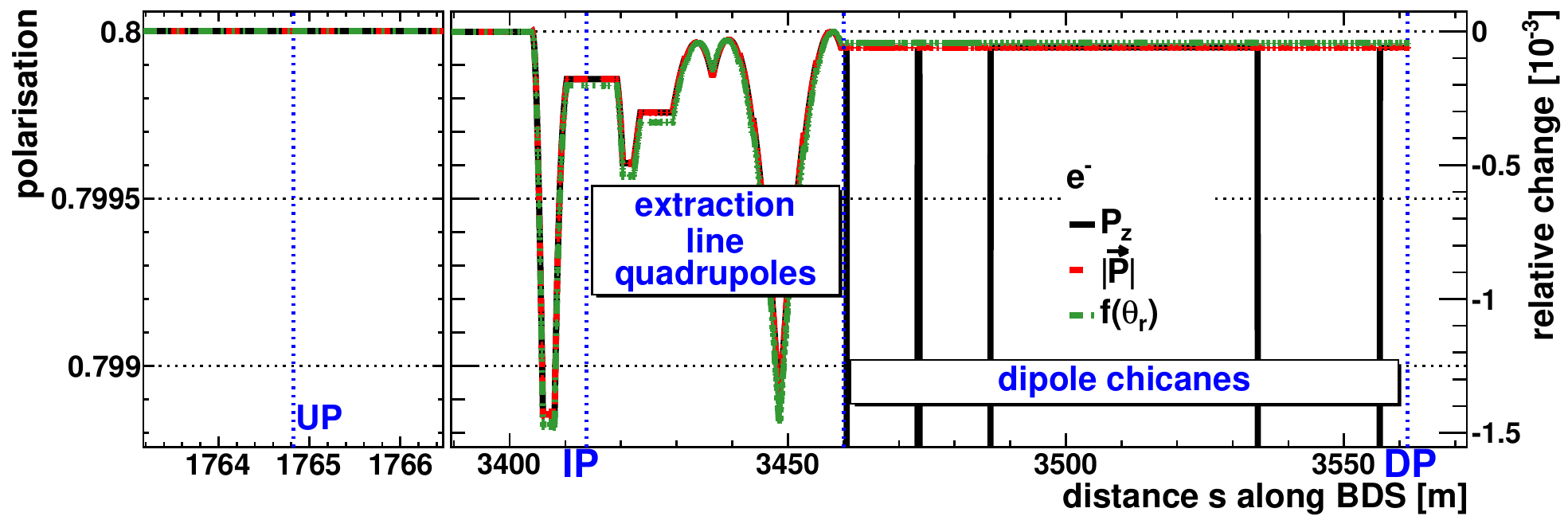}
   \caption{Longitudinal polarisation \polz\ (black, solid), magnitude of the polarisation vector \pol\
            (red, dashed) and the function $f(\thr)$ of the angular divergence defined in
            equation~\ref{eq:fThr} (green, dash-dotted) of the electron beam at the upstream polarimeter
            (UP, left part) and between $e^+e^-$ IP and downstream polarimeter (DP, right part).}
   \label{fig:PolSIdeal}
\end{figure}

Figure~\ref{fig:PolSIdeal} shows the polarisation \pol\ and its longitudinal component \polz\ of the electron
beam at the upstream polarimeter and between the $e^+e^-$ IP and the downstream polarimeter for perfect
alignment and in absence of detector magnets and crab cavities. It shows that the final focus magnets (in
front of the IP) and the extraction line quadrupoles affect \pol\ and \polz\ likewise, while the polarisation
vector precesses in the dipole chicanes, but returns to a longitudinal configuration behind a pair of dipoles
with opposite fields. The influence of the quadrupoles is well described by the spin fan-out prediction
$f(\thr)$ (equation~\ref{eq:fThr}) which is drawn as green dash-dotted line. The residual differences stem
most likely from the beam energy spread, which is not taken into account by $f(\thr)$, or from the betatron
oscillations.

\begin{table}[hbt]
   \renewcommand{\arraystretch}{1.05}
   \centering
   \begin{tabular}{llll@{\hspace{7mm}}l@{\hspace{7mm}}l@{\hspace{7mm}}l}
      Simulation      &              & \multicolumn{1}{c}{UP} & \multicolumn{2}{c}{$e^+e^-$ IP}     & \multicolumn{2}{c}{DP}\\
      \hline
      \thr            & [$\mu$rad]   & $0.986 \pm 0.004$      & \multicolumn{2}{l}{$35.4 \pm 0.1$}  & \multicolumn{2}{l}{$16.27 \pm 0.06$} \\
      $(0.8-\polz)$   & [$10^{-6}$]  & $0.2 \pm \varepsilon$  & \multicolumn{2}{l}{$143.0 \pm 0.8$} & \multicolumn{2}{l}{$48.4 \pm 0.3$}   \\
      $(0.8-\pol)$    & [$10^{-6}$]  & $0.2 \pm \varepsilon$  & \multicolumn{2}{l}{$143.0 \pm 0.8$} & \multicolumn{2}{l}{$48.4 \pm 0.3$}   \\
      $(0.8-f(\thr))$ & [$10^{-6}$]  & $0.1 \pm \varepsilon$  & \multicolumn{2}{l}{$162 \pm 1$}     & \multicolumn{2}{l}{$34.2 \pm 0.2$}   \\[3mm]
      \multicolumn{3}{l}{  }                                  & \multicolumn{2}{c}{$e^+e^-$ IP}     & \multicolumn{2}{c}{DP}               \\
      \multicolumn{3}{l}{Design values}                       & RDR       & TDR                     & RDR        & TDR                     \\
      \hline
      \thr            & \multicolumn{2}{l}{[$\mu$rad]}        & $35$      & $45$                    & $17$       &  $23$                   \\
      $(0.8-f(\thr))$ & \multicolumn{2}{l}{[$10^{-3}$]}       & $0.157$   & $0.25$                  & $0.039$    &  $0.064$                \\
   \end{tabular}
   \caption[]{Upper table: Angular divergence \thr, longitudinal polarisation \polz, magnitude of the
              polarisation vector \pol\ and the function $f(\thr)$ as defined in equation~\ref{eq:fThr}
              of the electron beam at the polarimeters (UP/DP = up-/downstream polarimeter) and the
              $e^+e^-$ IP. Lower table: \thr\ and $f(\thr)$ calculated from the design values
              (table~\ref{tab:beamParametersDesign}). The accuracy of the simulation is estimated to be
              $10^{-7}$ at best. Therefore, uncertainties below $0.5\cdot 10^{-7}$ are denoted as
              $\varepsilon$. Uncertainties between $0.5\cdot 10^{-7}$ and $1\cdot 10^{-7}$ are rounded up.}
   \label{tab:beamParameters}
\end{table}

Table~\ref{tab:beamParameters} lists selected bunch parameters from the simulation (and calculated from the
design values, as far as possible) at the polarimeters and at the $e^+e^-$ IP, where the latter serves
mainly for illustration here, since it does not affect the cross calibration of the polarimeters. In the
context of the envisaged precision of $0.1\,\%$, the simulation results are in good agreement with the
values predicted based on the angular divergence \thr. On this basis, the consequences of the stronger
focussing for the TDR parameters can be estimated as well. Assuming the spin fan-out at the upstream
polarimeter to be negligible, the relative difference in \polz\ at the two polarimeters due to spin
fan-out for TDR parameters amounts to $\Delta\polz/\polz=8.0\cdot 10^{-5}$. This is $64\,\%$ larger
than for RDR parameters, but still very small compared to the aim of $10^{-3}$.

We also studied the effect of a finite knowledge of the beam parameters, in particular the emittances.
A variation of $10\,\%$ leads to a negligible change in polarisation of $3\cdot 10^{-5}$.

The effects arising from the emission of synchrotron radiation in the beamline magnets have been
found to be negligible as well: The change in polarisation due to spin-flips between the two polarimeters
has been calculated to be $<10^{-6}$. The energy loss due to emission of synchrotron radiation and the
resulting changes in the particle trajectories change the polarisation by $5\cdot 10^{-6}$. For more
details, see section 9.1 of \cite{thesisMoritz}.

\subsection{Beam Alignment at the Polarimeters and at the $e^+e^-$ IP}  \label{subsec:beamAlignment}
In order to avoid measuring values for \polz\ at the polarimeters and at the $e^+e^-$ IP which differ
a priori due to T-BMT precession, there should be no relative angle between the orbits at these points.
Therefore, no effective bending angle between these points is foreseen in the BDS design.

Misalignments, however, might lead to relative incident angles $\Delta\thb$, which translates to a precession of the
polarisation vector by an angle of $\br{1+a\gamma}\cdot\Delta\thb$ (equation~\ref{eq:tbmtTheta}) if the energy spread can
be neglected \cite{Woods:2004qu}. The design requirement on the relative beam alignment between the polarimeter locations is
  $\Delta\thb \leq 50\,\mu$rad, driven by polarimetry demands.
To achieve the maximum longitudinal polarisation, spin rotators in front of the main linac are used
to adjust the polarisation vector parallel to the beam at the upstream polarimeter. This is expected
to be possible with an uncertainty of $\Delta\thp=25\,$mrad%
\footnote{A possible scheme to measure the alignment of the polarisation vector is presented in \cite{moenig}.}.

For a beam energy of $250$\,GeV, these two contributions give the following total uncertainty on the
polar angle of the polarisation vector:
\begin{equation}
   \Delta\thp^\text{tot} = \sqrt{ \Delta\thp^2 + \brbig{\br{1+a\gamma}\cdot \Delta\thb}^2} = 38\,\text{mrad}
\end{equation}
The corresponding uncertainty of the longitudinal polarisation amounts to
\begin{equation}
   \Delta \polz / \polz = 1-\cos\br{\Delta\thp^\text{tot}} = 0.72\cdot 10^{-3}.
   \label{eq:tabAngleCorrection1}
\end{equation}

The contribution from the incident angles increases with the beam energy. For a beam energy of $500\,$GeV
instead of $250\,$GeV, $\Delta\thp^\text{tot}$ rises to $62\,$mrad and $\Delta\polz/\polz$ to
$1.9\cdot 10^{-3}$. This is by far dominated by the contribution from the beam alignment: even for perfect
alignment of the polarisation vector, $\Delta\polz/\polz$ would still amount to $1.6\cdot 10^{-3}$.
Thus, an upgrade to a collision energy of $1\,$TeV would require an improved beam alignment in order
to achieve the precision goal of $0.1\,\%$ for the spin transport in the BDS.

A computational correction of the measured \polz\ for known incident angles would in principle be possible,
but gives an additional contribution to the uncertainty of \polz. As shown in section~7.4 of
\cite{thesisMoritz}, such a correction seems extremely difficult in view of the precision goal of $0.1\,\%$,
since the contribution to $\Delta \polz / \polz$ grows with the angle itself as well
as with its uncertainty. Therefore, the beam orbits at the polarimeters and at the $e^+e^-$ IP have to be
aligned by correction magnets. This correction method gives an additional contribution to the uncertainty
of \polz, which is discussed in the following.

\subsection{Residual Effects from the Beam Orbit Correction}       \label{subsec:orbitCorrectionEffects}
Misalignments of magnets as described in section~\ref{subsec:misalignments} lead to misalignments of the
beams. To keep the beams close to the design orbit and bring the bunches to collision at the $e^+e^-$ IP,
the BDS is equipped with a number of correction dipoles. However, these dipoles can not prevent the beam
from leaving the design orbit in the first place, but only bend it back to the design orbit. While it was
assumed in the previous section that a re-alignment of the beam orbit would also fully restore the
alignment of the polarisation vector, the accuracy of this assumption will be tested in this section.

\begin{table}[hb]
   \centering
   \begin{tabularx}{\textwidth}{lX}
      Sample name & Description                                                               \\
      \hline
      \samnom   & No misalignments (same data as for figure~\ref{fig:PolSIdeal}).             \\
      \sammfive & RMS size of all offsets  $5\,\mu$m, RMS size of all rotations  $5\,\mu$rad. \\
      \sammten  & RMS size of all offsets $10\,\mu$m, RMS size of all rotations $10\,\mu$rad. \\
 \end{tabularx}
   \caption[]{Simulated samples for the investigation of the effects of magnet misalignments.}
   \label{tab:dataSamplesMisalignments}
\end{table}

For this purpose, random misalignments of the beamline magnets, i.e.\ offsets and rotations in all three
dimensions have been introduced into the simulation. Three data samples with different RMS sizes of
misalignments have been produced as listed in table~\ref{tab:dataSamplesMisalignments}. For each of the
1000 bunches per sample, new misalignments are generated.
Correlations in space or time have not been taken into account for the misalignments.
To correct the beam orbit, the corrector magnets foreseen in the lattice are used. The magnet strengths
are computed from the beam positions at the beam position monitors (BPMs)~\cite{Boogert:2013sea} such that
the beam position offsets measured by the BPMs are minimised. For the current study, negligible BPM
resolutions have been assumed.
At the $e^+e^-$ IP, the fast-feedback correction is emulated which adjusts the beam position and incident
angle according to the foreseen tolerances. The correction procedure is described in detail
in \cite{thesisMoritz}.

\begin{figure}[ht]
   \includegraphics[width=\columnwidth]{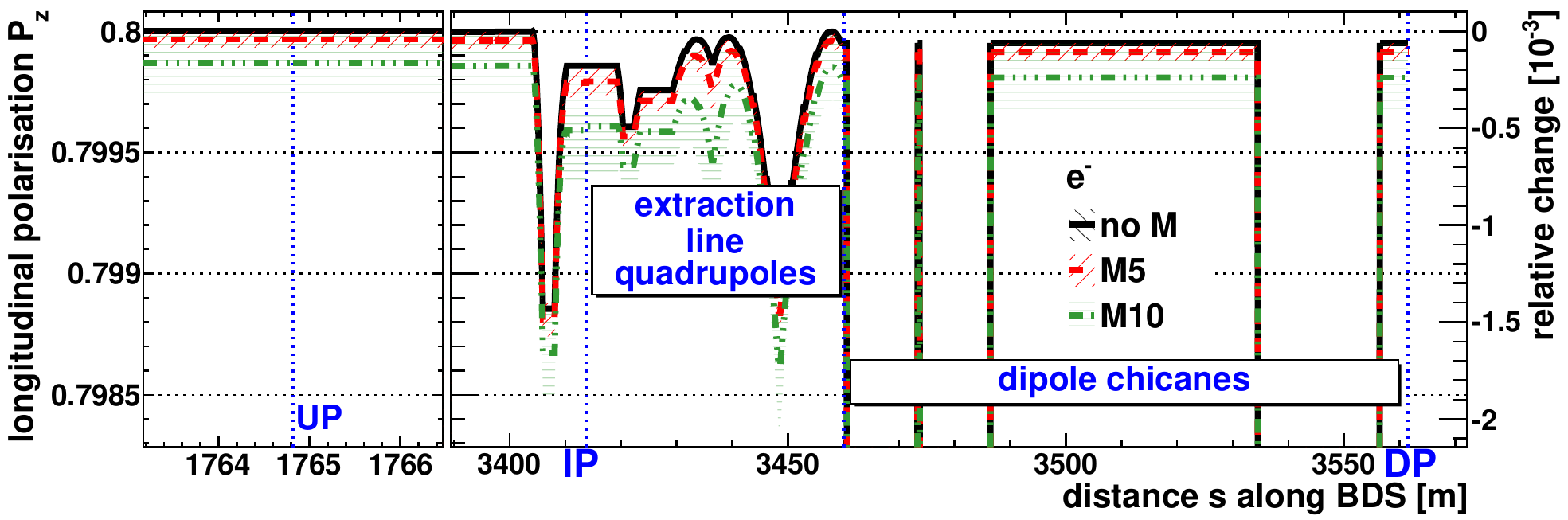}
   \caption{Longitudinal polarisation \polz\ of the electron beam at the upstream polarimeter (UP, left part) and between $e^+e^-$ IP and downstream polarimeter (DP, right part) for the data samples listed in table~\ref{tab:dataSamplesMisalignments}. The uncertainty bands correspond to the RMS spread of all runs. For the perfectly aligned case, the band is too small to be visible.}
   \label{fig:PolSIdealMisal}
\end{figure}

\begin{table}[hbt]
   \renewcommand{\arraystretch}{1.05}
   \centering
   \begin{tabular}{l|lll@{\hspace{5mm}}l@{\hspace{5mm}}l}
                  &                 &              & \multicolumn{1}{c}{\samnom} & \multicolumn{1}{c}{\sammfive} & \multicolumn{1}{c}{\sammten} \\
      \hline
      Upstream    & \thb            & [$\mu$rad]   & $0.006 \pm 0.004$           & $2 \pm 1$                     & $3 \pm 2$                    \\
      polarimeter & \thr            & [$\mu$rad]   & $0.986 \pm 0.004$           & $0.989 \pm 0.005$             & $1.00 \pm 0.01$              \\[2mm]
                  & $(0.8-\polz)$   & [$10^{-6}$]  & $0.2 \pm \varepsilon$       & $34 \pm 34$                   & $132 \pm 128$                \\
                  & $(0.8-\pol)$    & [$10^{-6}$]  & $0.2 \pm 0.1$               & $0.2 \pm \varepsilon$         & $0.3 \pm 0.1$                \\
                  & $(0.8-f(\thr))$ & [$10^{-6}$]  & $0.1 \pm \varepsilon$       & $0.1 \pm \varepsilon$         & $0.1 \pm \varepsilon$        \\
      \hline
      Downstream  & \thb            & [$\mu$rad]   & $0.21 \pm 0.08$             & $0.7 \pm 0.4$                 & $1.2 \pm 0.6$                \\
      polarimeter & \thr            & [$\mu$rad]   & $16.27 \pm 0.06$            & $16.4 \pm 0.4$                & $16.7 \pm 0.9$               \\[2mm]
                  & $(0.8-\polz)$   & [$10^{-6}$]  & $48.4 \pm 0.3$              & $86 \pm 37$                   & $192 \pm 138$                \\
                  & $(0.8-\pol)$    & [$10^{-6}$]  & $48.4 \pm 0.3$              & $49 \pm 2$                    & $50 \pm 5$                   \\
                  & $(0.8-f(\thr))$ & [$10^{-6}$]  & $34.2 \pm 0.2$              & $35 \pm 2$                    & $36 \pm 4$                   \\
   \end{tabular}
   \caption[]{Incident angle \thb, angular divergence \thr, longitudinal polarisation \polz, magnitude of
              the polarisation vector \pol\ and the function $f(\thr)$ as defined in equation~\ref{eq:fThr}
              of the electron beam at the polarimeters for the data samples listed in
              table~\ref{tab:dataSamplesMisalignments}. For $\varepsilon$, see the explanation in
              the caption of table~\ref{tab:beamParameters}}
   \label{tab:beamParametersMisalignments}
\end{table}

Figure~\ref{fig:PolSIdealMisal} shows the longitudinal polarisation \polz\ at the upstream polarimeter and
between the $e^+e^-$ IP for the same data samples. In addition, table~\ref{tab:beamParametersMisalignments}
lists selected bunch parameters of the electron beam at the polarimeters for different sizes of misalignments
for the data samples listed in table~\ref{tab:dataSamplesMisalignments}.
The uncertainties in the table and the shaded areas in the figure denote the RMS spread of \polz\ of
the simulated bunches. These spreads reflect the possible variations depending on
the exact misalignments of the individual magnets, which lead to different trajectories for each bunch.

While the norm of the polarisation vector is hardly affected by the misalignments, there is a larger
effect on the longitudinal component \polz. The values of \pol\ remain consistent with $f(\thr)$
(cf.\ section~\ref{subsec:spinTransport}); thus, the behaviour of \pol\ is still fully explained
by spin fan-out.
The increases in the mean value and the uncertainty of \thb\ do, however, not suffice to explain the
decrease in \polz\ by precession of the polarisation vector according to equation~\ref{eq:tbmtTheta}:
$\Delta\polz=2\cdot 10^{-4}$ would correspond to $\thb=39\,\mu$rad, while $\thb=5\,\mu$rad would yield
only $\Delta\polz=4\cdot 10^{-6}$. This discrepancy appears after one of the correction magnets in the
``dogleg'' where the electron beam by-passes the photon target area of the positron source (cf.\ fig.~\ref{fig:bds-SB2009}), possibly due to non-commutation of rotations about different axes: while all deflections of the bunch by misaligned magnets are compensated for with the corresponding deflections at the correction magnets, this compensation does not work perfectly for the polarisation vector which performs rotations by the $(1+a\gamma)$-fold deflection angles.

\begin{figure}[ht]
   \centerline{\includegraphics[width=0.5\columnwidth]{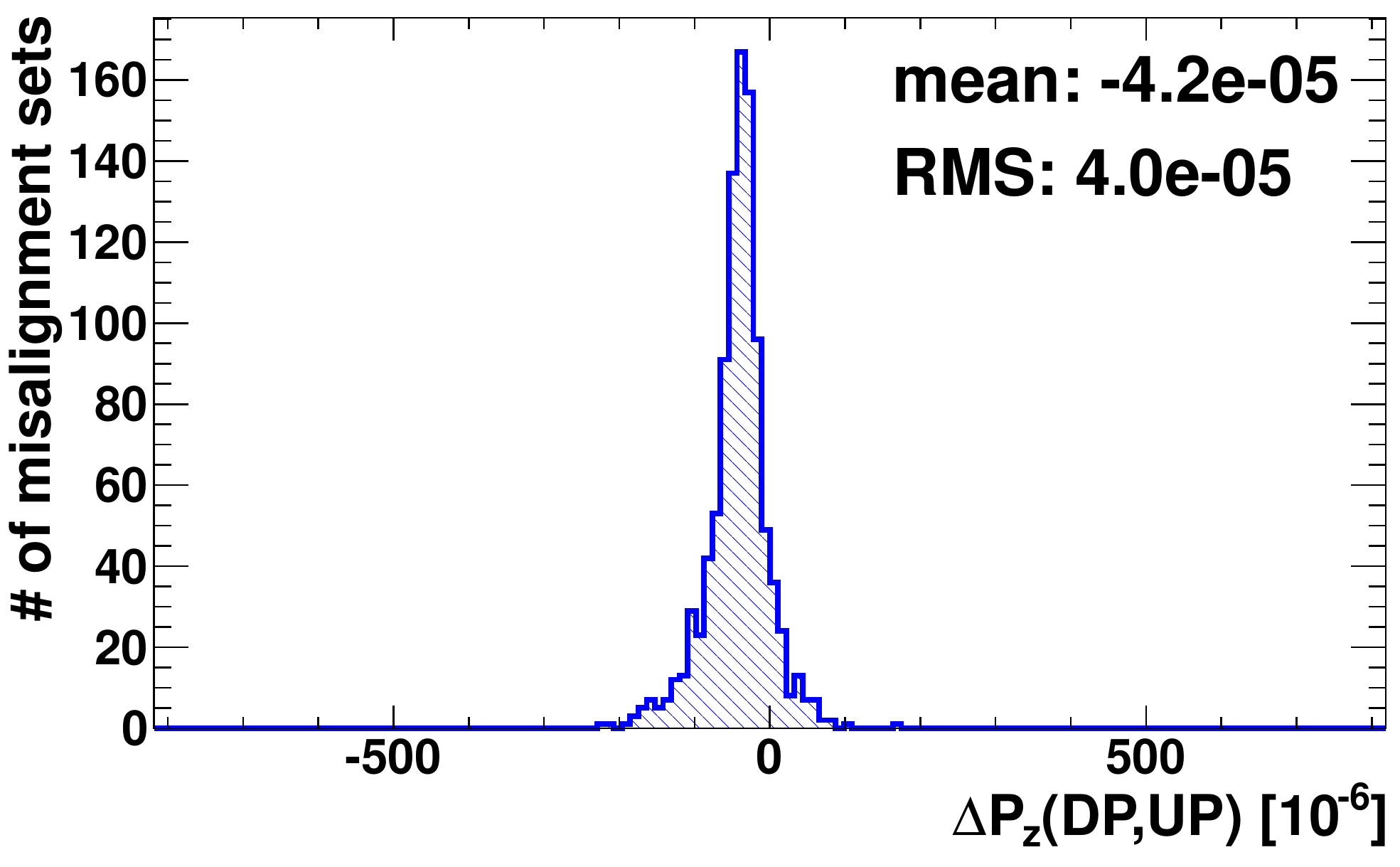}}
   \caption{Difference of the longitudinal polarisation \polz\ of the electron beam at the upstream polarimeter (UP) and the downstream polarimeter (DP) for the misalignment sample ``\sammten''.}
   \label{fig:ZpolDPvsUPEM10}
\end{figure}

In case of purely static misalignments, the decisive quantity for the cross calibration of the polarimeters
is the residual difference of the longitudinal polarisation at both locations. Figure~\ref{fig:ZpolDPvsUPEM10} shows this difference in terms of $\Delta\polz({\mathrm{UP,DP}}):= (\polz^{{\mathrm{UP}}}-\polz^{{\mathrm{DP}}})/\sqrt{2}$
for the sample  ``\sammten''. It follows an approximately Gaussian distribution with an RMS spread of $0.04\cdot 10^{-3}$. This is about a factor $10$ smaller than the variations of \polz\ at any individual location, i.e. the uncertainty band in figure~\ref{fig:PolSIdealMisal}.
Nevertheless, we conservatively estimate the uncertainty from the magnet misalignments and the beam orbit correction
based on figure~\ref{fig:PolSIdealMisal}, in order to allow for
 time-dependent misalignments which have not been explicitely studied here.
Thus, we define this uncertainty as the relative deviation of the lower bound of the spread for \polz\ for the samples with misalignments
from the mean value of \polz\ for the sample without misalignments (both values at
the downstream polarimeter, neglecting the decrease in \polz\ at the upstream polarimeter).
This yields a relative uncertainty of $0.09\cdot 10^{-3}$ for the sample ``\sammfive'' and
$0.35\cdot 10^{-3}$ for the sample ``\sammten''.

Thus, we conclude that misalignments of beamline elements are not the dominating source of
systematic uncertainty, but still can give a sizable contribution.
Especially the impact of time-dependent effects should be studied in more detail based on
realistic ground-motion models of the selected ILC site, also taking into account finite BPM resolutions.

\subsection{Detector Magnets and Crab Cavities}       \label{subsec:commissioning}
In this section we investigate more closely the interaction region devices and their impact on
the spin transport. When operating the machine without collisions, the detector magnets and the
crab cavities could be switched on or off. Thus, their impact on the polarisation measured at the
downstream polarimeter could rather easily be disentangled from instrumental differences between the
up- and downstream polarimeter measurements.

\subsubsection{Detector Magnets}       \label{subsubsec:detectorMagnets}
The beams enter the detector under an angle of 7\,mrad, which is half the beam-beam crossing angle.
Hence, both the main solenoid and the anti-DID feature magnetic field components parallel and
perpendicular to the beam.

The parallel field components might destroy the synchronisation between the angles of the beam orbit
and the polarisation vector, such that \polz\ takes different values at the $e^+e^-$ IP and at the
polarimeters even if the beam orbit is aligned parallel at these three locations. This effect has,
however, turned out to be negligible ($\Delta\polz\sim 10^{-5}$) \cite{thesisMoritz} despite the
strong longitudinal field of the solenoid. This can be understood since the transverse components
of the polarisation vector, which would undergo the precession about the longitudinal axis, are rather
small, and since the precession about the longitudinal axis does not speed up with rising particle
energies (cf.\ equation~\ref{eq:tbmtOmegaBdecomposed}). The effect of the longitudinal field components
would have to be re-evaluated for physics operation of the ILC with transverse beam polarisation.

The perpendicular field components act like dipole magnets on the beam. The resulting deflection would
alter the beam position by several micrometers at the $e^+e^-$ IP, such that no collisions would occur,
and several millimeters at the downstream polarimeter. To correct the beam position, additional dipole
magnets in front of and behind the detector are foreseen, which are not contained in the lattice files
\cite{SB2009lattice} yet. For establishing luminosity, only the relative angles of the beams to each
other are relevant \cite{antididAndrei}. However, polarimetry requires also the angles of the beam
orbits at the polarimeters to be aligned parallel to the beam orbits at the $e^+e^-$ IP to a level of
$50\,\mu$rad (cf.\ section~\ref{subsec:beamAlignment}) in presence of collisions. Without any beam
orbit correction, the fields of the detector magnets would induce a deflection angle of
$\thb=185\,\mu$rad at the $e^+e^-$ IP, which leads to a decrease in longitudinal polarisation by
$\Delta\polz=4.5\cdot 10^{-3}$ due to T-BMT precession \cite{thesisMoritz}. As explained in the
previous section, such a deflection angle at the $e^+e^-$ IP can only be corrected for if it can be
measured to significantly better than $10\,\%$. If for instance the above $\thb=185\,\mu$rad are known
to $15\,\mu$rad ($8\,\%$), the resulting uncertainty on the polarisation from this correction alone
amounts to $0.1\,\%$. Thus, it is very important to foresee sufficient orbit correction possibilities
such that not only the relative, but also the absolute angles of the beams at the $e^+e^-$ IP can be
adjusted in presence of the detector magnets.

\subsubsection{Crab Cavities}       \label{subsubsec:crabCavities}
To initiate the rotation of the particle bunches, the crab cavities generate a time-dependent
electromagnetic field. In the centre of the cavity and at the time of the bunch passages, this field
can be approximated by a time-dependent magnetic dipole field, which deflects the bunch particles
horizontally by an angle proportional to the longitudinal distance from the bunch centre. This deflection
goes along with the corresponding spin precession (equation~\ref{eq:tbmtTheta}), which results in a spin
fan-out along the $z$-axis and thus in a decrease of the polarisation. The effect on the polarisation
can, however, be expected to be small, since the fan-out vanishes at the bunch centre, where the majority
of the particles is located. Simulations have shown that the effect on the polarisation is smaller than
$10^{-5}$ (section 8.1 in \cite{thesisMoritz}).

\subsection{Cross Calibration of the Polarimeters}       \label{subsec:summaryCrossCalib}
Table~\ref{tab:uncertainties} summarises the uncertainties on the spin transport between the polarimeters
for the cross calibration in absence of collisions and for a beam energy of $250\,$GeV. The major
contribution comes from the alignment precision of the beam and the polarisation vector
(section~\ref{subsec:beamAlignment}). A second large contribution comes from the beam orbit correction
(section~\ref{subsec:orbitCorrectionEffects}), while all other contributions listed are negligible.
All contributions sum up to an uncertainty of $0.080\,\%$, which matches the goal of at most $0.1\,\%$.
For higher beam energies, many of the contributions can be expected to increase. The contribution from
the alignment precision alone rises to $1.9\cdot 10^{-3}$ for a beam energy of $500\,$GeV, which
necessitates a better beam alignment.

\begin{table}[hbt]
  \renewcommand{\arraystretch}{1.05}
   \centering
   \begin{tabularx}{\textwidth}{Xr}
      Contribution                                                            & $\delta\polz/\polz\ [10^{-3}]$ \\
      \hline
      Beam and polarisation alignment at polarimeters                         & $0.72$                         \\
      (assuming $\Delta\thb=50\,\mu$rad, $\Delta\thp=25\,$mrad) &                                \\
      Random misalignments ($10\,\mu$m$/\mu$rad) with beam orbit correction   & $0.35$                         \\
      Variation in beam parameters ($10\,\%$ in the emittances)               & $0.03$                         \\
      Longitudinal precession in detector magnets                             & $0.01$                         \\
      Bunch rotation to compensate the beam crossing angle                    & $<0.01$                        \\
      Emission of synchrotron radiation                                       & $0.005$                        \\
      \hline
      Total                                                                   & $0.80$                         \\
   \end{tabularx}
   \caption{Contributions to the uncertainty of the spin transport from the upstream to the downstream polarimeter for a beam energy of $250$\,GeV in the absence of collisions.
   \label{tab:uncertainties}}
\end{table}

\section{Luminosity-Weighted Average Polarisation}       \label{sec:lumipol}
In this section, we discuss the relevant quantity for the interpretation of collision data,
i.e.\ the luminosity-weighted average polarisation at the IP. We studied the direct depolarisation
in collisions as well as for the first time the effects which arise
at the downstream polarimeter location and their interplay with polarimetry. Finally, we
lay out a strategy how to access the luminosity-weighted average polarisation from polarimeter
measurements and how to compare this to collision data.

In addition to the RDR beam parameter set used in section~\ref{sec:crosscal}, also collisions with
TDR beam parameters have been simulated (cf.\ section~\ref{subsec:ilcbds}). As explained in
section~\ref{sec:crosscal}, the current lattice features a final focus according to RDR parameters
at the $e^+e^-$ IP.
However, no significant changes in the spin tracking upstream of the $e^+e^-$ IP are expected.
On the other hand, the extraction line lattice is suitable for both parameter sets. Thus, the beams are
generated at the $e^+e^-$ IP, passed to the simulation of the collisions, and then propagated to
the downstream polarimeter.

These two data samples named ``\samrdr'' and ``\samtdr'' are both generated with the spin configuration
as described in the beginning of section~\ref{sec:crosscal}.
This is well motivated since the spin transport is dominated by spin fan-out as shown in the previous
section. To illustrate nevertheless the effects of a different initial spin configuration, a third
sample ``\samtdrstar'' has been generated:
all macroparticle
polarisation vectors are aligned along the beam axis at the $e^+e^-$ IP, instead of being fanned out
according to the focussing of the beam. Such a different spin configuration could e.g.\ be the result
of the betatron oscillations (non-commutating rotations, cf.\ section~\ref{subsec:orbitCorrectionEffects}).

To be able to distinguish the contributions from spin precession and from energy losses, each of the
three data samples has been resimulated with the particle energies fixed to $E=250\,$GeV
(\samrdrfixedEnergy, \samtdrfixedEnergy, \samtdrstarfixedEnergy). In other words, the bunches are
generated without initial energy spread, and no energy loss due to synchrotron radiation or beamstrahlung
(and thus no depolarisation due to Sokolov-Ternov effects) is simulated, so that spin precession
is the only source of changes in polarisation. The definitions of all samples are summarised in
table~\ref{tab:dataSamples}.

\begin{table}[hb]
   \centering
   \begin{tabularx}{\textwidth}{lX}
      Sample name & Description    \\
      \hline
      \samrdr     & Beam parameters according to the RDR (cf.\ table~\ref{tab:beamParametersDesign}),
                    but with the energy spread according to the TDR (same parameters as in
                    section~\ref{sec:crosscal}).\\
      \samtdr     & Beam parameters according to the TDR (cf.\ table~\ref{tab:beamParametersDesign}).\\
      \samtdrstar & Like TDR, but with a different initial spin configuration: all macroparticle
                    polarisation vectors are aligned along the $z$-axis at the IP, instead of being
                    fanned out according to the focussing of the beam.\\
      \samrdrfixedEnergy, \samtdrfixedEnergy, \samtdrstarfixedEnergy & Samples like above, but all
                    particles with fixed energy $E=250\,$GeV (no energy spread, no synchrotron
                    radiation or beamstrahlung).\\
 \end{tabularx}
   \caption[]{Simulated samples for the investigation of the beam-beam collision effects.}
   \label{tab:dataSamples}
\end{table}

The luminosities of the simulated collisions amount to $(2.02\pm 0.02)\cdot 10^{34}$\,cm$^{-2}$\,s$^{-1}$
and $(1.52\pm 0.03)\cdot 10^{34}$\,cm$^{-2}$\,s$^{-1}$ for the samples \samrdr\ and \samtdr, respectively,
which is in good agreement with the design luminosities (cf.\ table~\ref{tab:beamParametersDesign}).

\subsection{Beam Properties at Downstream Polarimeter}     \label{subsec:measurablePolarisation}
In absence of collisions, the bunches at the Compton IPs of the polarimeters are smaller than the laser
spots or of similar size. In collision mode, the beams get disrupted by the beam-beam interaction.
Although the downstream polarimeter is placed at a secondary focus of the extraction line optics,
the refocussing is hampered by the larger emittance and the larger energy spread of the disrupted beam
after the collision.

\begin{figure}[htb]
   \begin{center}
      \begin{subfigure}{.475\linewidth}
         \includegraphics[width=\textwidth]{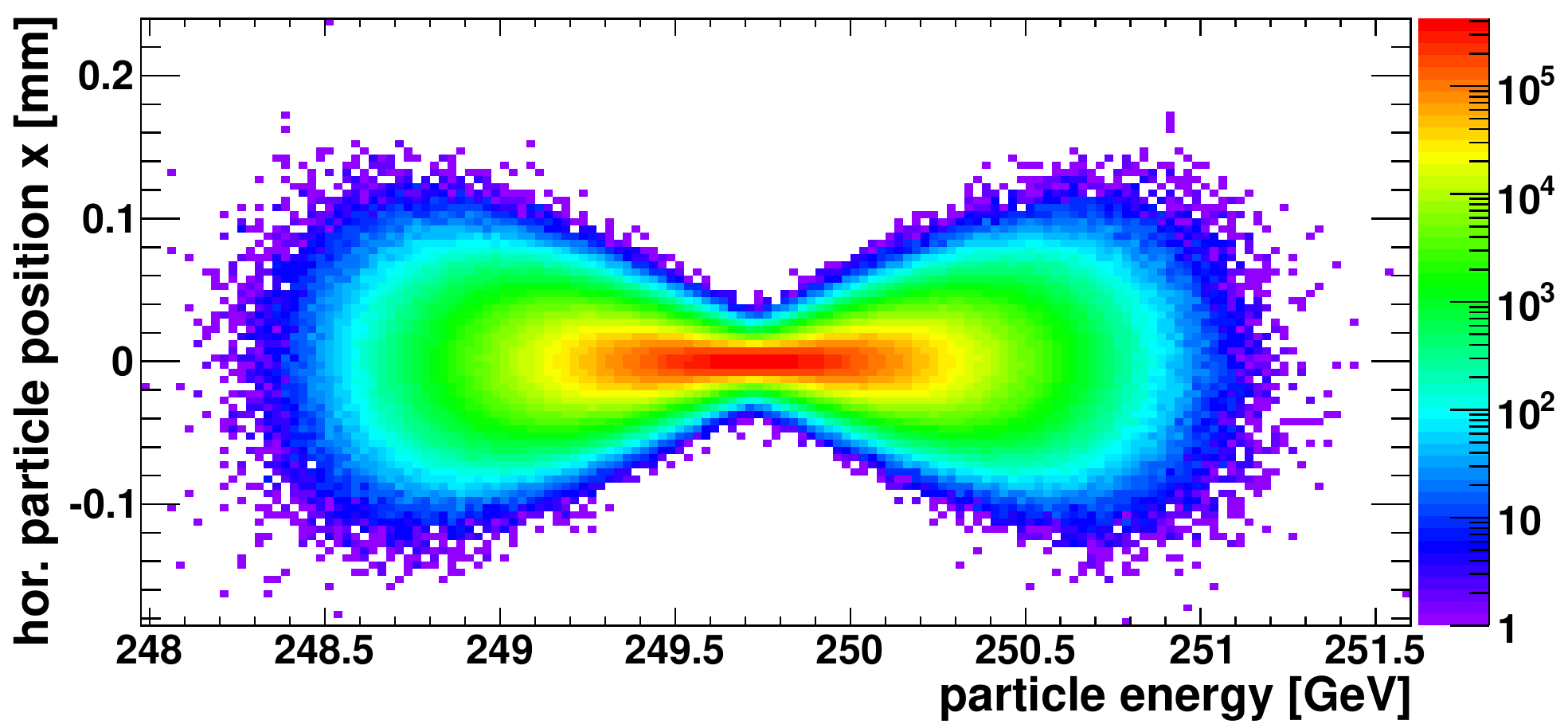}
         \subcaption{} 
      \end{subfigure}
      \begin{subfigure}{.475\linewidth}
         \includegraphics[width=\textwidth]{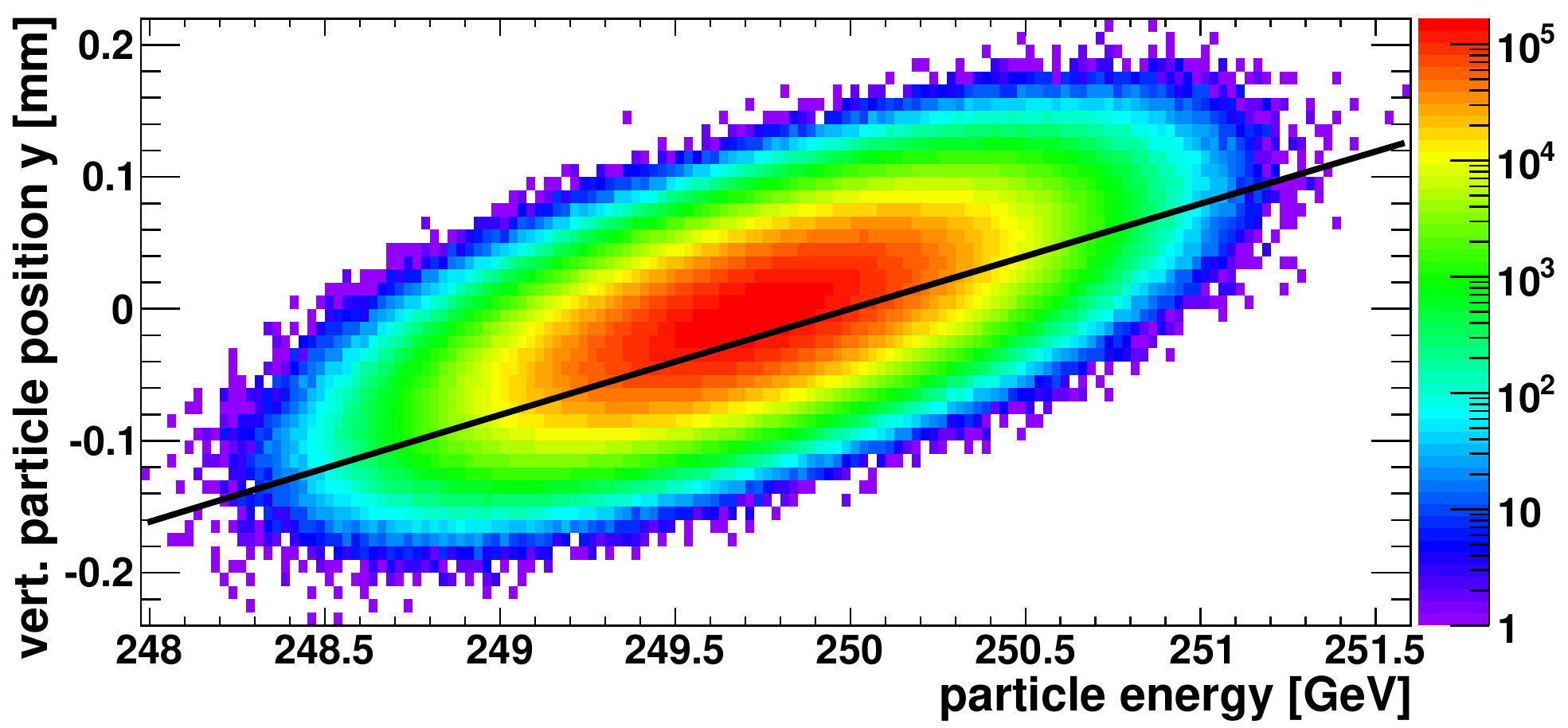}
         \subcaption{} 
      \end{subfigure}
      \ \\
      \begin{subfigure}{.475\linewidth}
         \includegraphics[width=\textwidth]{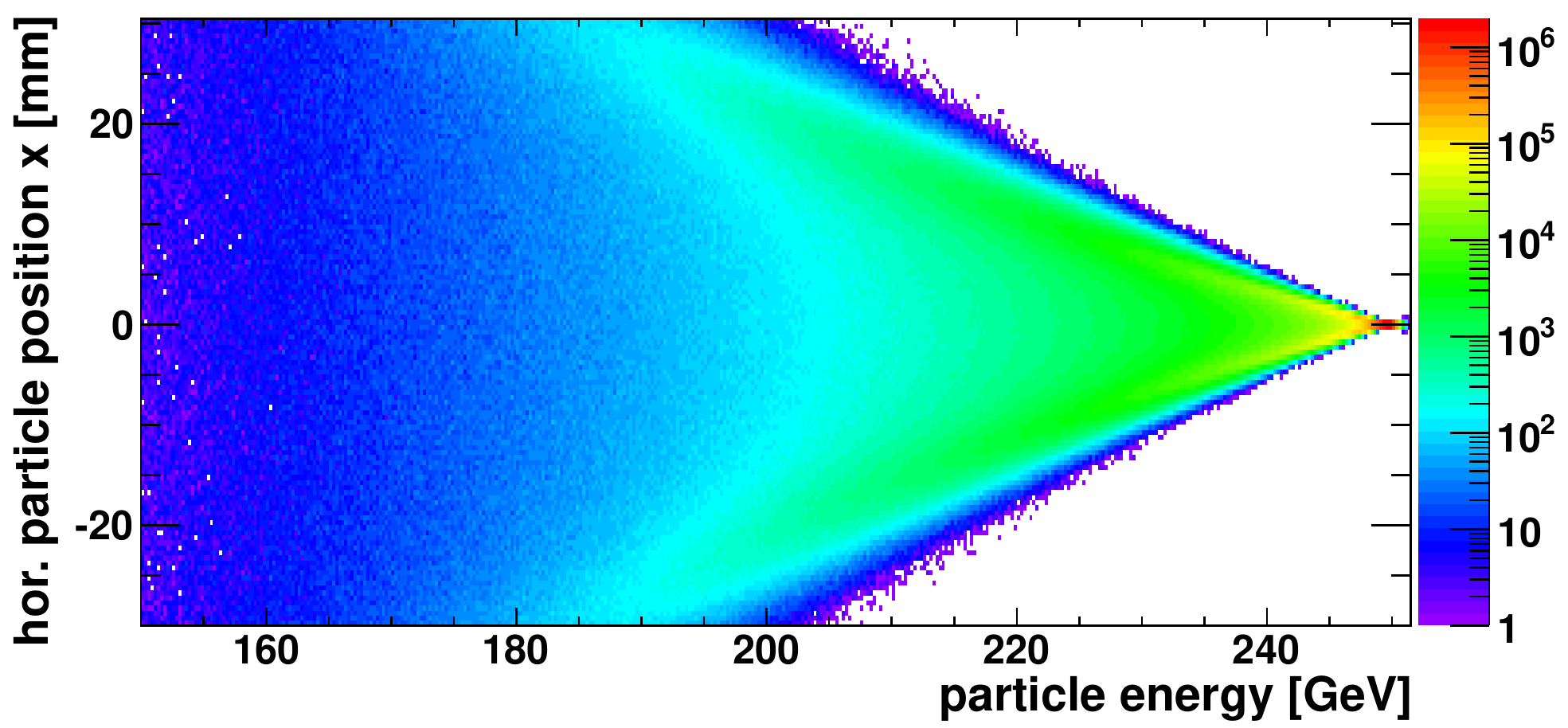}
         \subcaption{} 
      \end{subfigure}
      \begin{subfigure}{.475\linewidth}
         \includegraphics[width=\textwidth]{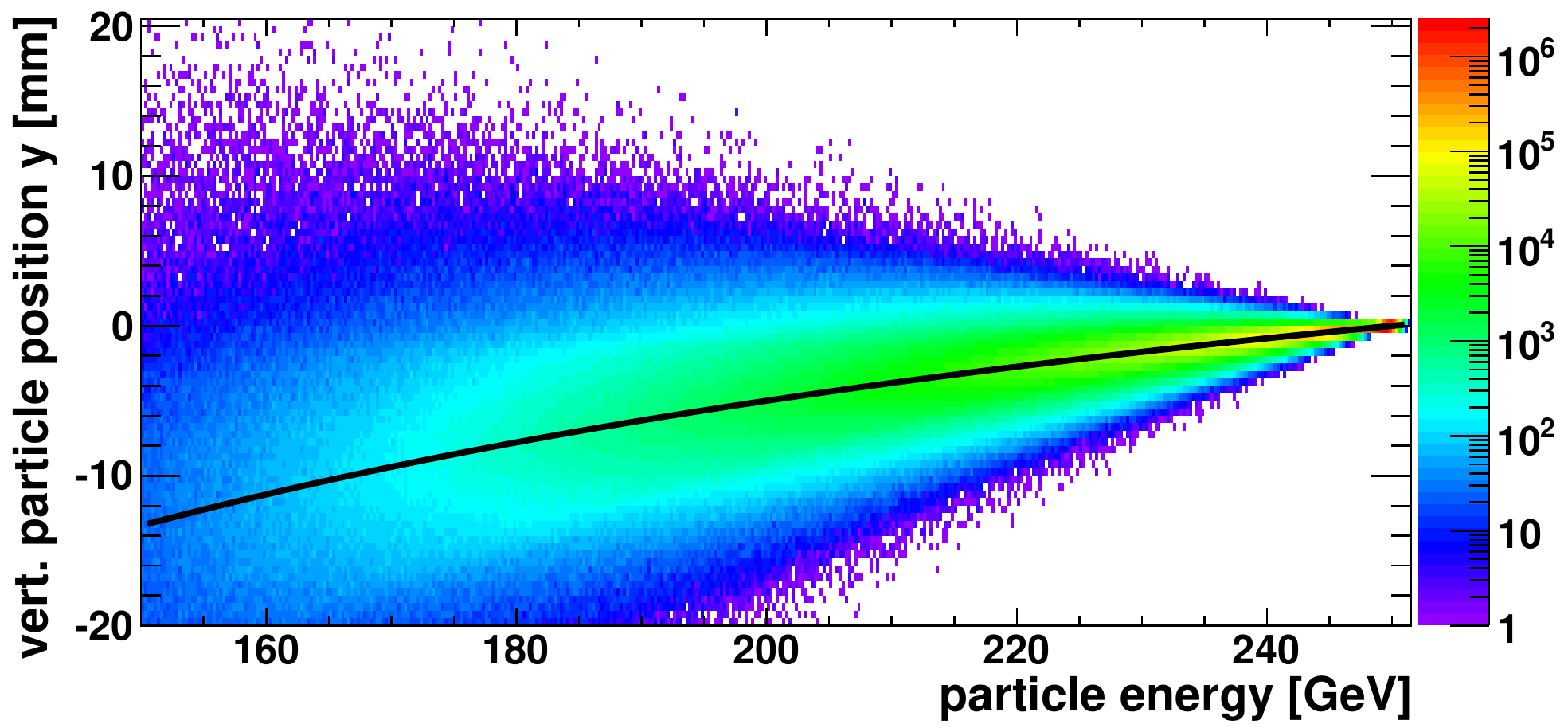}
         \subcaption{} \label{fig:dispersionyDPSubfigB}
      \end{subfigure}
      \ \\
      \begin{subfigure}{.475\linewidth}
         \includegraphics[width=\textwidth]{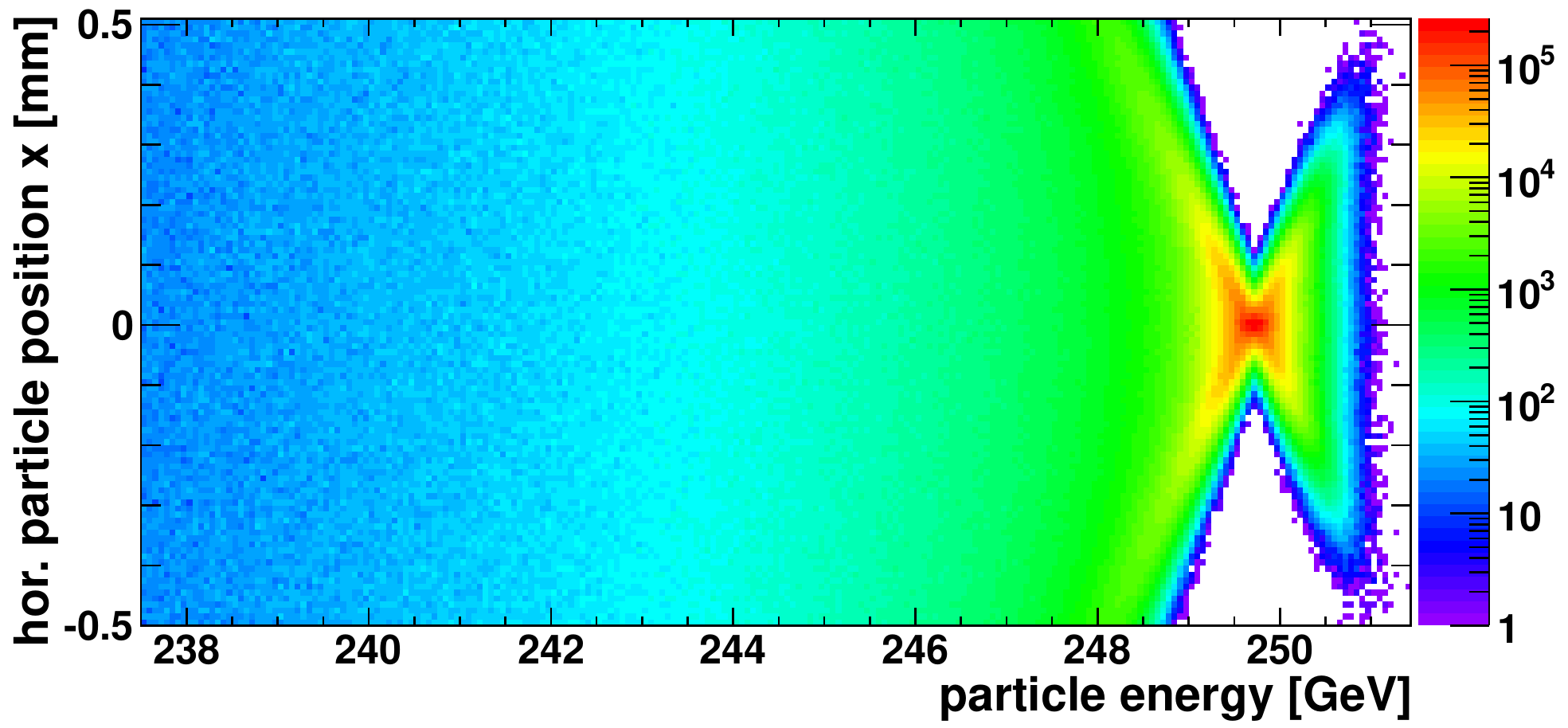}
         \subcaption{}
      \end{subfigure}
      \begin{subfigure}{.475\linewidth}
         \includegraphics[width=\textwidth]{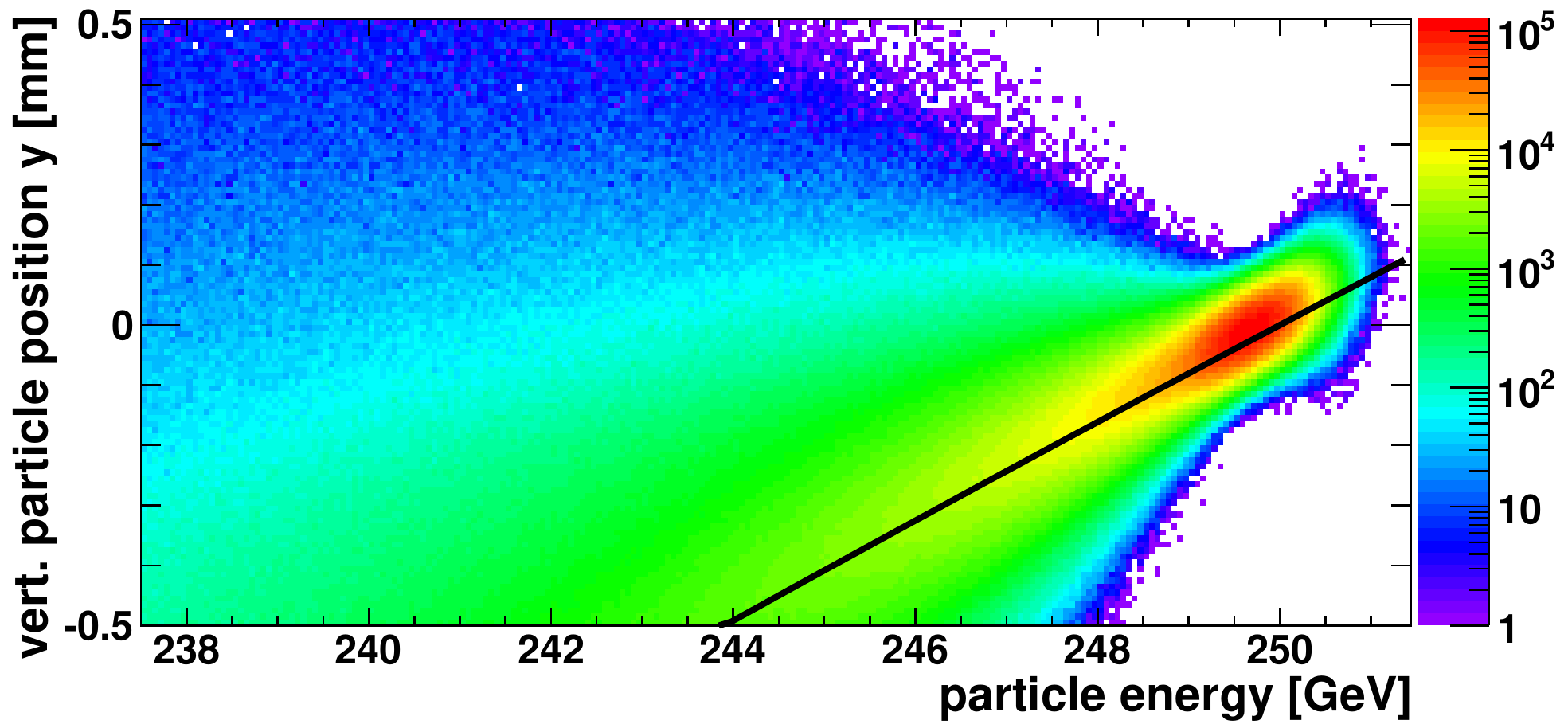}
         \subcaption{} \label{fig:dispersionyDPSubfigC}
      \end{subfigure}
   \end{center}
   \caption{Horizontal and vertical particle positions vs the energy at the downstream polarimeter
            for 1000 electron bunches without collisions (a,b), after (c,d) the collision. (e,f) show zooms into (c,d).
            The black solid curve in (b,d,f) shows the dispersion relation between $y$-position and energy.
   }
   \label{fig:dispersionxDP}
\end{figure}

This is illustrated by figure~\ref{fig:dispersionxDP}, which shows the horizontal and vertical
particle positions vs. the particle energies for the sample \samtdr\ at the downstream polarimeter
without collisions (a,b) and after a collision (c-f).
Since the downstream polarimeter is located in a vertical magnet chicane
(figure~\ref{fig:downstreamPolChicane}), dispersion also contributes to the vertical beam size.
The expected offset due to dispersion is indicated as a black line for comparison.
In absence of collisions, the particle bunch sizes are well matched by the design laser spot size
of $\sigma_{x\gamma}=\sigma_{y\gamma}=100\,\mu$m (cf.\ section~\ref{sec:polarimetry}). In contrast,
the disrupted beam after a collision extends over centimeters, although a large fraction of particles
has lost no or only little energy and is still well focussed.
These particles are confined to a spot of $\sim 100\,\mu$m size, while the particles which have
lost more energy are spread out much further and are not covered by the laser spot of the polarimeter.

\begin{figure}[hbt]
   \begin{center}
      \begin{subfigure}{.475\linewidth}
         \includegraphics[width=\textwidth]{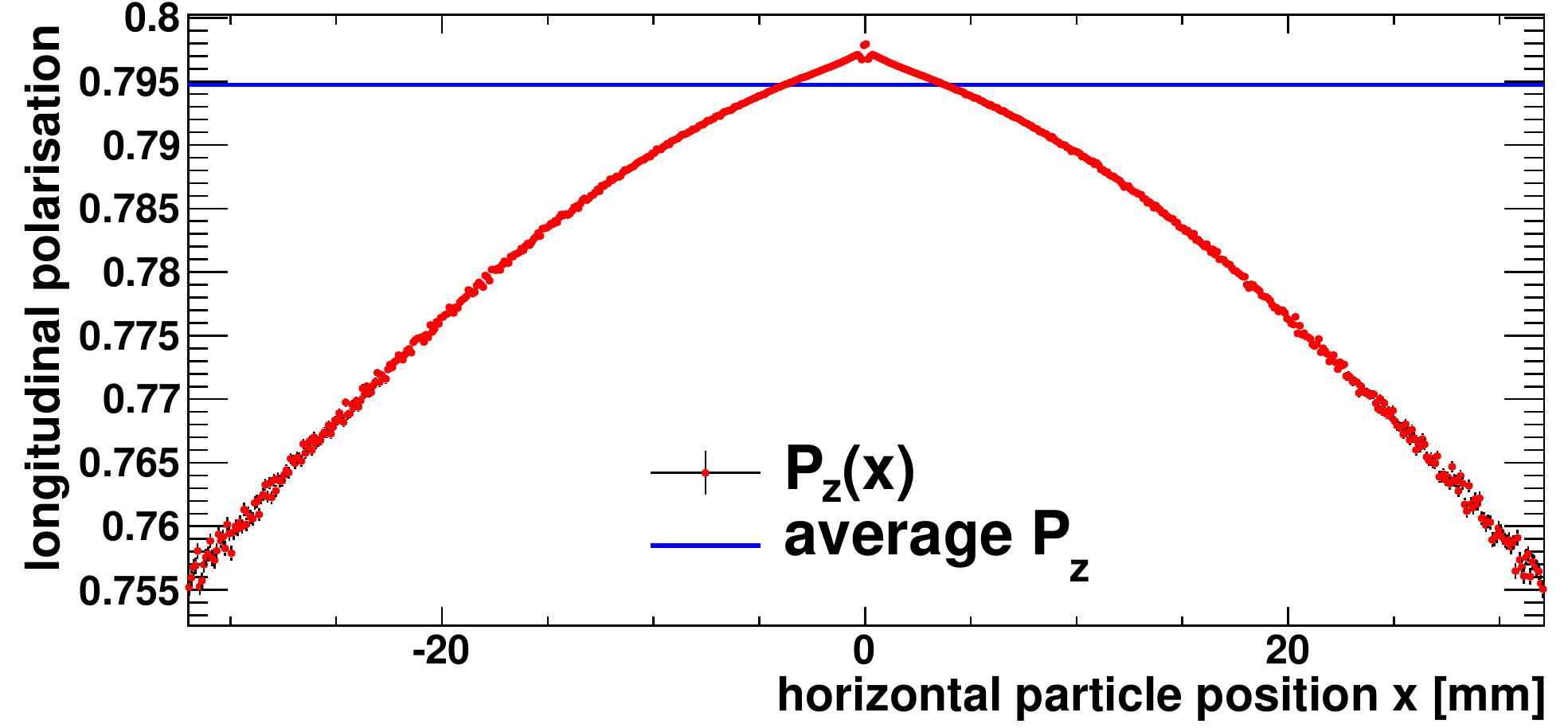}
         \subcaption{} \label{fig:corrPolEYDPSubfigA}
      \end{subfigure}
      \begin{subfigure}{.475\linewidth}
         \includegraphics[width=\textwidth]{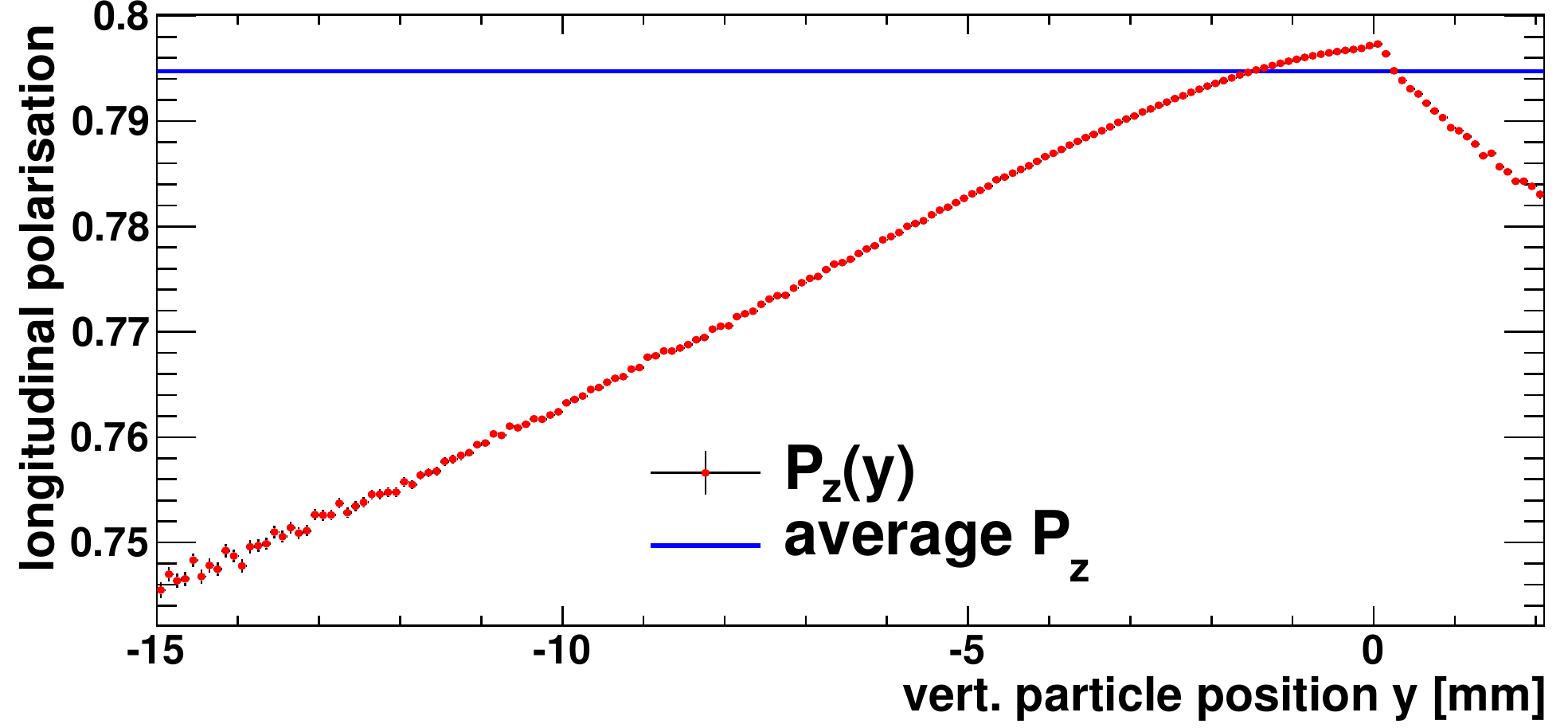}
         \subcaption{} \label{fig:corrPolEYDPSubfigB}
      \end{subfigure}
     \ \\
      \begin{subfigure}{.475\linewidth}
         \includegraphics[width=\textwidth]{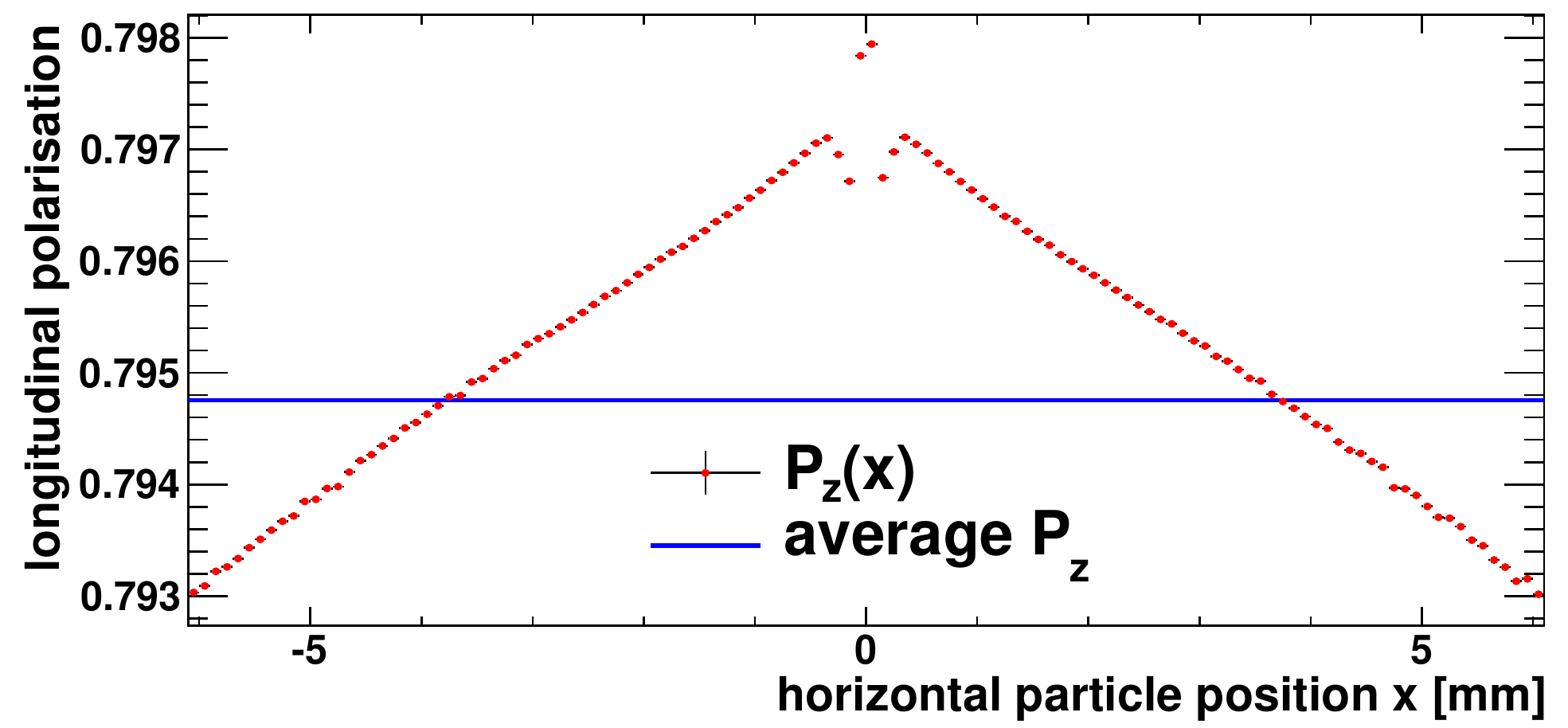}
         \subcaption{} \label{fig:corrPolEYDPSubfigC}
      \end{subfigure}
      \begin{subfigure}{.475\linewidth}
         \includegraphics[width=\textwidth]{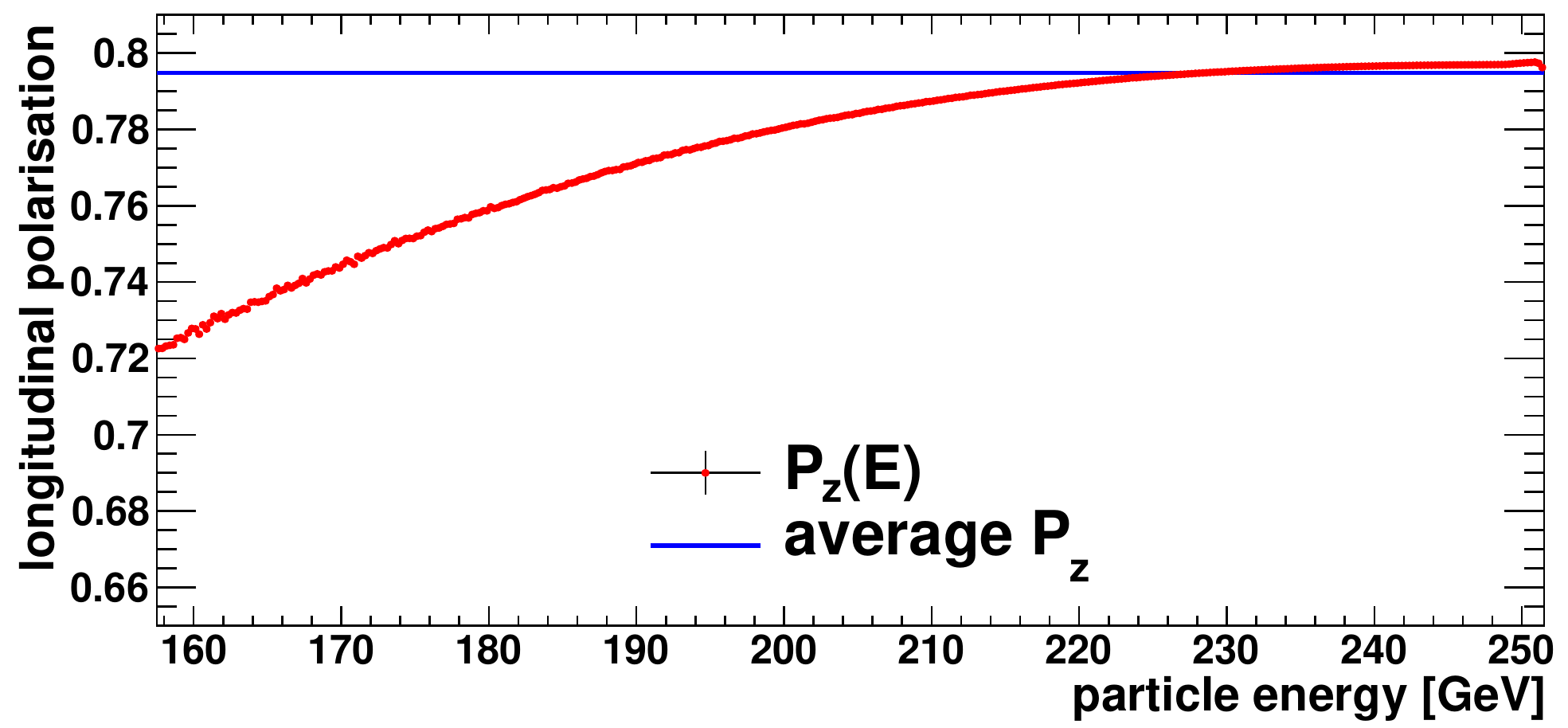}
         \subcaption{} \label{fig:corrPolEYDPSubfigD}
      \end{subfigure}
   \end{center}
   \caption[]{Longitudinal polarisation of the macroparticles versus the horizontal particle position
              $x$ (a) (and zoom (c)), the verticle particle position $y$ (b) and the particle energy $E$ (d)
              at the downstream polarimeter.
   }
   \label{fig:corrPolEYDP}
\end{figure}

As illustrated by figure~\ref{fig:corrPolEYDP}, the longitudinal polarisation in the electron beam
(sample \samtdr, after collision) correlates strongly with the position (a-c) and the energy (d) of the
macroparticles, since particles which have been deflected more strongly by electromagnetic fields during
the collision experience more spin precession and emit more beamstrahlung.

For a measurement using a laser spot of $\sim 100\,\mu$m size, this implies that the measurement will
not only be rather sensitive to the correct positioning of the laser, but also that the outcome of
the measurement depends on the exact size of the laser spot at the Compton IP. As a first estimate of
the effect on the downstream polarisation measurement, we investigated the \textit{measurable polarisation},
which we define as the average longitudinal polarisation of the subset of macroparticles which would
be hit by the laser, i.e.\ within a slice through the bunch defined by the laser spot size and
the vertical crossing angle.

\subsection{Collision Effects on Polarisation and Polarimetry}
As discussed in section~\ref{subsec:beambeamCollisions}, the luminosity-weighted average polarisation
during collisions can be restored at the location of the downstream polarimeter if the extraction line
quadrupoles halve the divergence angle compared to its value at the $e^+e^-$ IP. This behaviour, however,
is only obtained if spin fan-out is the dominating effect on the polarisation,
which should be the case for the samples with pure T-BMT precession, \samrdrfixedEnergy,
\samtdrfixedEnergy\ and \samtdrstarfixedEnergy.

\begin{figure}[hbt]
   \begin{center}
      \includegraphics[width=0.75\textwidth]{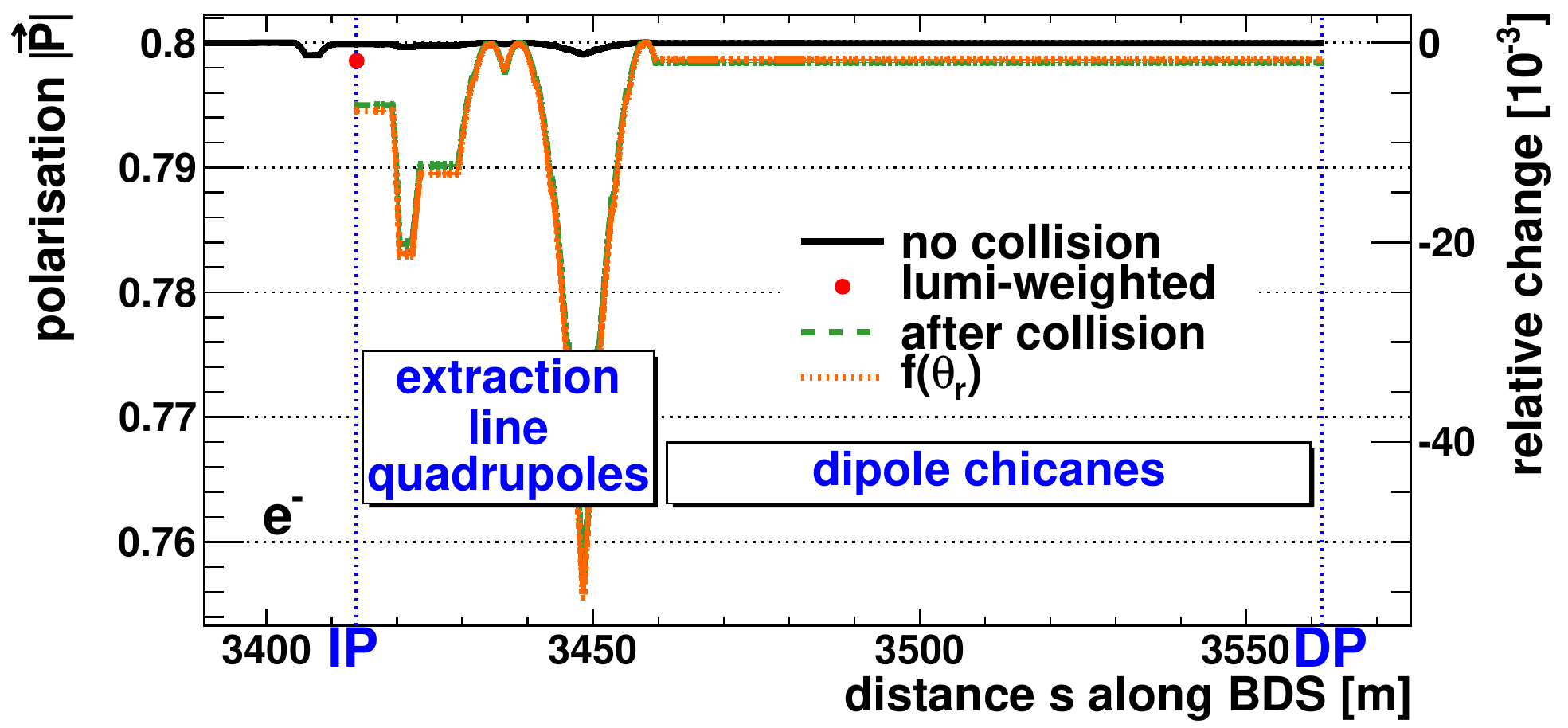}
   \end{center}
   \caption[]{Magnitude of the polarisation vector of the electron beam along the BDS in the absence
              of any energy loss (data sample \samrdrfixedEnergy). Shown are the propagation in the
              case of no collisions, the different propagation after the collision, the luminosity-weighted
              polarisation in the collision and $f(\thr)$ after the collision.}
   \label{fig:PolSCollNodE}
\end{figure}

Figure~\ref{fig:PolSCollNodE} shows the polarisation \pol\ of the electron beam between the IP and the
downstream polarimeter for the case of \samrdrfixedEnergy\ after the collision and without collisions.
Due to the disruption in the collision, the divergence angle of the beam is enlarged from $33\,\mu$rad
to $205\,\mu$rad at the IP, which leads to a decrease in polarisation of $0.5\,\%$ by spin fan-out.
The luminosity-weighted polarisation in the collision, indicated by the red dot at the IP, is lower
than the incoming polarisation by $0.27\,\%$, which is in accordance with equation~\ref{eq:yokoyaChen16}.
Behind the IP, the full tracking result is in perfect agreement with the expectation from pure
spin precession, $f(\thr)$. At the location of the downstream polarimeter, the divergence angle
is halved to $102\,\mu$rad as desired. Thus, the value of the luminosity-weighted polarisation
is reproduced at the downstream polarimeter to a level of $0.01\,\%$ despite spin fan-out of
several percent along the way. This larger spin fan-out occurs since the disrupted beam is more divergent
and hence refocussed more strongly than a non-colliding beam.
In summary, the design concept for the spin transport in the extraction line works perfectly if only
T-BMT precession is taken into account, but no energy losses or radiative depolarisation.

\begin{figure}[hbt]
   \begin{center}
      \begin{subfigure}{0.75\linewidth}
         \includegraphics[width=\textwidth]{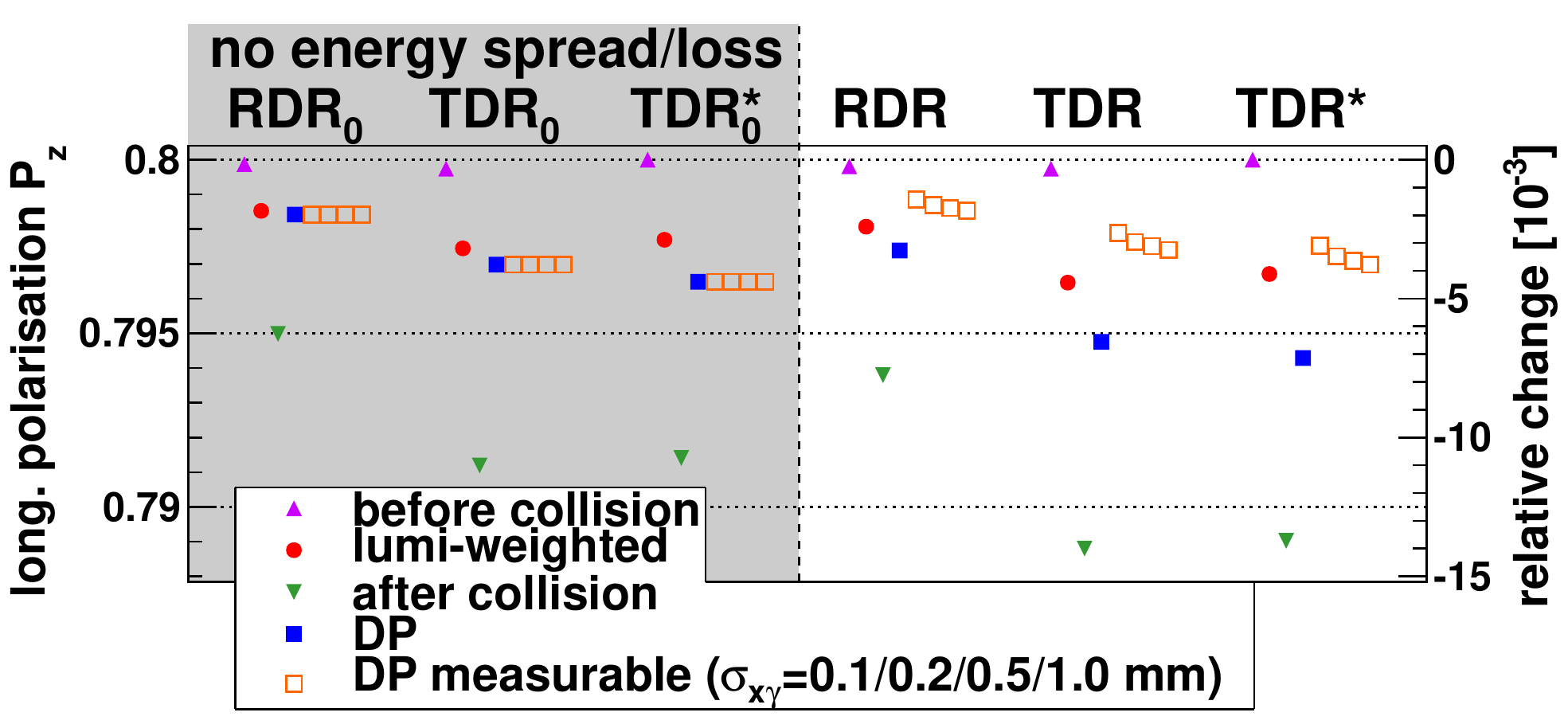}
         \subcaption{} \label{fig:PolOverview7SubfigA}
      \end{subfigure}
       \ \\
      \begin{subfigure}{0.75\linewidth}
         \includegraphics[width=\textwidth]{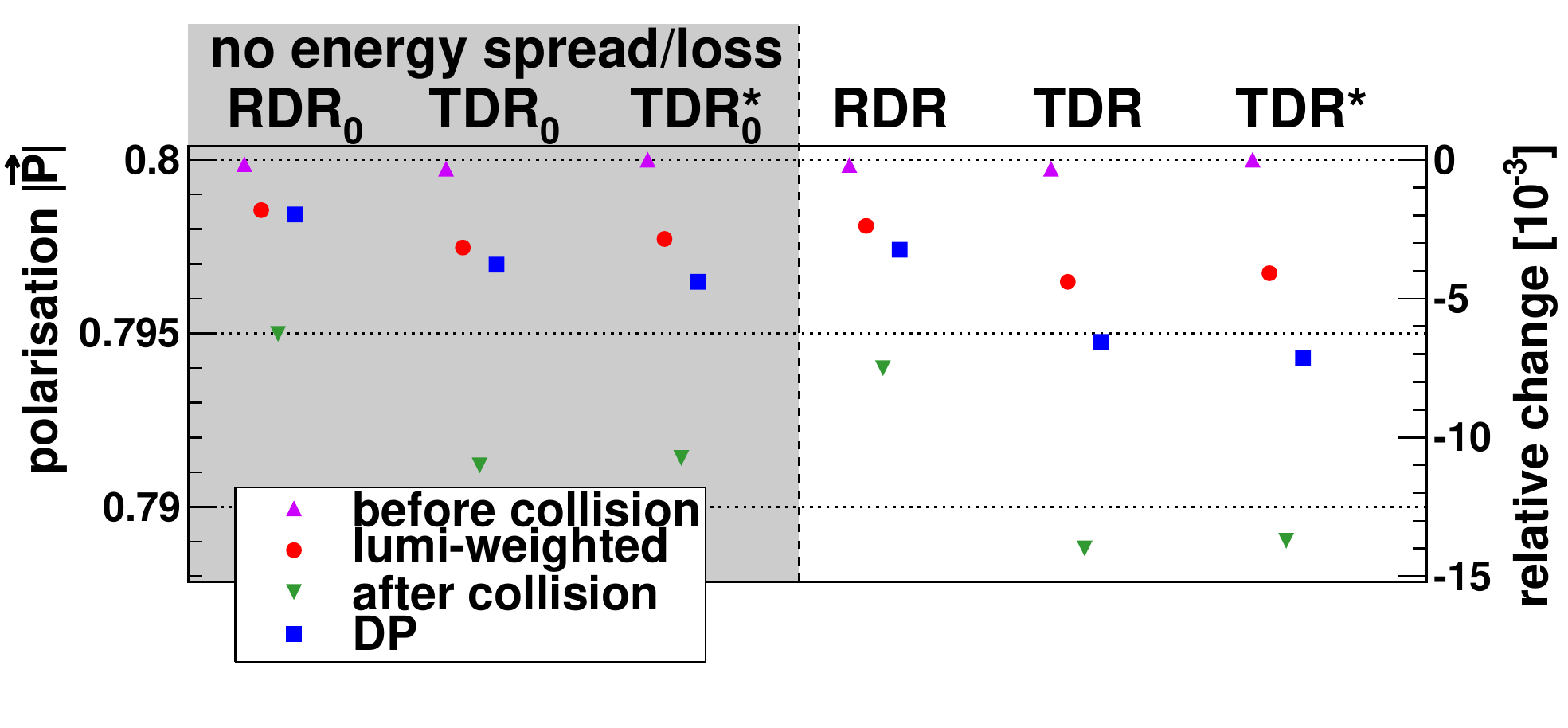}
         \subcaption{} \label{fig:PolOverview2SubfigB}
      \end{subfigure}
   \end{center}
   \caption{Electron beam polarisation along the BDS at the IP and at the polarimeters. Shown are
            the longitudinal polarisation \polz\ (a) and the magnitude of the polarisation vector
            \pol\ (b) of the electron beam at the IP before and after the collision as well as
            the luminosity-weighted values and the values at the downstream polarimeter (DP).
            In (a), also the measurable longitudinal polarisation for different laser spot sizes
            $\sigma_{x\gamma}$ ($=\sigma_{y\gamma}$) is shown.}
   \label{fig:PolOverview7}
\end{figure}

In reality, however, energy loss and radiative depolarisation are non-negligible and need to be included
in any realistic study. Therefore, all six
cases have been simulated in analogy to figure~\ref{fig:PolSCollNodE}, and the results
are summarised in figure~\ref{fig:PolOverview7}, by displaying the obtained polarisation
values only at the points of interest. For the longitudinal polarisation, the
{\itshape measurable polarisation} is indicated in addition for four different assumptions on the laser
spot size $\sigma_{x\gamma}$ ($=\sigma_{y\gamma}$), always assuming a perfect centering of laser
and electron beam.

The first column, \samrdrfixedEnergy , corresponds directly to figure~\ref{fig:PolSCollNodE},
showing again the perfect recovery of the luminosity-weighted average polarisation at the
downstream polarimeter. Since the electron beam spot is fully hit by the laser, the
{\itshape measurable polarisation} is identical to the longitudinal polarisation of the full bunch.

The next two columns show the results for TDR parameters, still for T-BMT precession only, but two
different start configurations of the spins. Interestingly enough, the luminosity-weighted average
polarisation and the polarisation at the downstream polarimeter
start to deviate by $0.06$ - $0.15\,\%$ even without energy loss, both in magnitude and in longitudinal
component. This could be due to the larger horizontal disruption parameter of the TDR parameter set
($D_x=0.3$ compared to $D_x=0.17$ in the RDR case), since equation~\ref{eq:yokoyaChen16} assumes
the horizontal disruption parameter to be small ($D_x \ll 1$). Moreover, the size of this deviation
depends on the initial spin configuration.
This indicates that
a full start-to-end spin tracking simulation from the electron and positron sources is
highly desirable for the future.

The realistic results including energy loss and Sokolov-Ternov depolarisation are shown in
the remainder of the figure. Compared to the corresponding sample without energy loss,
the depolarisation in collisions increases in all three cases as expected. More importantly,
also the differences between the polarisation at the downstream polarimeter and the luminosity-weighted
average one increase to tolerable $0.07\,\%$ in the \samrdr\ case, while it amounts to $0.17$ - $0.24\,\%$
in the \samtdr\ case, which exceeds the aim of $0.1\,\%$.

However, most strikingly the energy loss limits the refocussing of the beams, resulting in significantly
increased beam sizes at the location of the downstream polarimeter
discussed in the previous section. Now, the {\itshape measurable polarisation} deviates
significantly from the total average at the downstream polarimeter by up to $0.3\,\%$, and changes with
the assumed size of the laser spot by up to $0.07\,\%$. The difference between
the measureable polarisation and the luminosity-weighted average at the IP is somewhat
smaller due to a partial cancellation. In particular in the nominal \samtdr\ case, the net difference
amounts to $0.15\,\%$ for a nominal laser spot size of $100\,\mu$m. This number depends significantly
on the actual beam parameters at the IP: for the \samrdr\ case, the
analoguous difference is only $0.08\,\%$.

For the positron beam, a different behaviour could be expected due to the different degree of polarisation,
the different initial beam energy spread and the absence of the positron production system,
which only exists in the electron beamline. However, the most dominant effect, T-BMT precession,
scales linearly with \pol. Thus, the relative changes in the (longitudinal) polarisation of
the positron beam differ only slightly from those of the electron beam \cite{thesisMoritz}.

Not yet included in the above numbers are additional contributions from non-perfect centering of
the laser beam onto the electron beam nor any effect of the energy and position distributions of
the incoming electrons on the analysing power of the downstream polarimeter,
which needs to be addressed in future studies. The luminosity-weighted average polarisations above
are expressed in the laboratory system, already accounting for the crossing angle of the beamlines,
but not boosted to the centre-of-mass system.

\subsection{Accessing the Luminosity-Weighted Average Polarisation}
In view of the above results, assuming the downstream polarimeter to directly measure the
luminosity-weighted average polarisation at the electron-positron interaction point \polzll\ seems
sufficient for percent-level, but not for permille-level precision. Instead, a more sophisticated
strategy based on a detailed understanding of the collisions as a function of time is suggested
in the following.

\begin{enumerate}
\item{\bf Cross-calibration of the polarimeters:} The very first step, to be repeated regularly,
e.g.\ during maintainance of the main detectors, is the cross-calibration of the polarimeters without
collisions. The beam time requirement is given by the downstream polarimeter, which
reaches a statistical precision of below $0.1\,\%$ after about one hour\footnote{It should be
investigated in the future whether in absence of collisions the downstream measurement could collect
sufficient statistics on similar timescales as the upstream polarimeter.}. Systematic effects on the spin transport
are expected to be $<0.1\,\%$, provided the
orbit alignment goals at the two polarimeter locations can be reached.
\item{\bf Upstream polarimeter:} During collisions, the upstream polarimeter is essential to determine
the initial polarisation value, which is expected to be $0.25\,\%$ (RDR) to $0.4\,\%$ (TDR) above \polzll,
and to track time variations. Both aspects are needed to predict the expectation at the
downstream polarimeter, which measures just one (or a few) bunches out of each train.
\item{\bf Knowledge of collision properties:} In order to predict a) the depolarisation in collisions
and b) the beam properties at the downstream polarimeter, all possible means to
monitor the beam parameters before, during, and after collisions should be employed.
This includes the energy spectrometers~\cite{boogert}, the monitoring of the instantaneous luminosity,
the beamstrahlung and the pair background in the forward calorimeters of
the main detectors~\cite{Abramowicz:2010bg} and the Gamma Calorimeter in the downstream polarimeter chicane.
From their combined information, the beam parameters, in particular $\sigma_{xe}$ and $\sigma_{ye}$
can be determined to about $10\,\%$~\cite{Grah:2008zz}. For comparison, the main difference between the
RDR and TDR beam parameter sets studied here is a reduction of $\sigma_{xe}$ by $25\,\%$. Thus
we estimate that the depolarisation in collisions could be predicted to about $0.1\,\%$ based on
the measured beam parameters.
\item{\bf Collision simulation:} Based on theoretical understanding of the intense-field QED environment
of the collisions, the knowledge of initial beam parameters and the post-collision diagnostics,
simulations can be employed to predict \polzll\ itself, as well as the energy and position distributions
and the polarisation of the spent beam at the downstream polarimeter location. Redundancy in instrumentation
is a key to validate the simulation tools.
\item{\bf Downstream polarimeter:} The prediction of the beam properties
at the downstream polarimeter location then needs to be employed to determine the changes in analysing power
compared to the well defined situation without collisions. Furthermore, the effect of uncertainties
in the laser alignment and focussing need to be assessed. The possibility to defocus the laser on purpose
and/or scan the outer regions of the electron beam should be considered in order to ``map out''
the spent beam and verify the effects predicted by simulations.
\item {\bf Comparison of polarimeters and simulations:} Based on the previous steps,
the measured polarisation value at the downstream polarimeter can be extrapolated to
the luminosity-weighted average at the IP. This needs to happen in a time-dependent manner,
since the intensities of the bunch-bunch collisions might change during a bunch train and on
longer time scales. This value can then be compared to the prediction for \polzll\ based on the
upstream polarimeter measurement and the beam parameter estimates. Only if sufficient agreement
is found in this step, permille-level accuracy on \polzll\ can be claimed.
It should be noted that once agreement is established, the extrapolated upstream polarimeter measurement
will be the main ``working horse'' due to its ability to simultanously measure
the polarisation for each bunch position in a train, while for the downstream polarimeter it takes
order of days to sample each bunch position with permille-level statistical precision.
\item {\bf Polarisation values for physics analyses:} The relevant effective polarisation
for the physics process under study (cf.\ section~\ref{subsec:beambeamCollisions}) then needs
to be calculated from values of \polzll\ obtained for both beams from the beam-parameter-dependent
extrapolation of the polarimeter measurements. Finally a luminosity-weighted average of the effective
polarisations can be formed for each particular data set, including
e.g.\ the analysis-specific run selection based on sub-detector availabilities. Here it should be noted
that instrumental uncertainties of the polarimeters as well as effects of misalignments should be
uncorrelated between both beams, while in particular the collision
effects are due to the mutual influence of the beams on each other and thus are correlated to a large extent.
\item {\bf Comparison with collision data:} After accumulation of a significant amount of collision data,
the long-term average of \polzll\ can be extracted also from the collision data themselves.
Typically several $100$\,fb$^{-1}$ distributed over all four helicity configurations are needed
for sub-percent statistical precision~\cite{Moenig, thesisIvan, EPSprocedings}, where again
time-dependencies and correlations can only be resolved based on
polarimeter information. Nevertheless, polarisation-sensitive Standard Model processes with
sufficiently large cross sections will provide an essential verification of the absolute
polarisation scale.
\end{enumerate}

\section{Conclusions}      \label{sec:conclusions}
The polarimetery concept for the ILC is based on the combination of two complementary Laser-Compton
polarimeters per beam and the long-term average polarisation determined
from collision data. These three ingredients can only be exploited coherently in conjunction
with detailed simulations of the spin transport and the beam-beam collision effects. For this purpose,
the simulation framework {\tt STALC} has been developed. It is not intrinsically
limited to the ILC, but can be run on any lattice and beam parameter set.

In absence of collisions, various effects which could influence the spin transport between
the polarimeters and, thus,
e.g.\ their cross-calibration have been evaluated. The dominating uncertainty stems from
the relative beam alignment at the two polarimeter locations. Here, the design goal of
$\Delta \thb \leq 50\,\mu$rad is just sufficient for permille-level precision for a beam energy
of $250$\,GeV, while for the upgrade to $\sqrt{s}=1$\,TeV an improvement to about
$\Delta \thb \sim 25\,\mu$rad would be required in order to maintain the same level of
precision on the polarisation.

In presence of collisions, additional effects have to be considered with respect to the polarisation.
This includes both the depolarisation in collision and thus the luminosity-weighted average polarisation
during the collision as well as the properties of the spent beam which influence
the downstream polarimeter measurement. It has been shown that reducing the angular divergence at
the downstream polarimeter with respect to the IP by a factor of two can only partially restore the
luminosity-weighted average polarisation at the downstream polarimeter location, since
for ILC beam parameters the energy loss and Sokolov-Ternov spin flips due to the emission
of beamstrahlung cannot be fully neglected anymore. The influence of the spent beam properties
on the downstream polarimeter measurement itself have been estimated roughly by
taking into account the finite laser spot size and the crossing angle. The variations found
are again at a level relevant for permille-level precision. Thus, it will be crucial to monitor
the collision parameters in real-time at least to the $10\,\%$-level by the forward calorimeters
of the main detectors and additional diagnostics like the GamCal and the extraction-line energy spectrometer.
With this knowledge, the luminosity-weighted average
polarisation at the IP can be extracted from both the upstream and the downstream polarimeter measurements
with complementary systematic uncertainties.

In the future, the influence of the spent beam properties on the downstream polarimeter measurement
as well as the possibilities to realise the long-range orbit alignment to the tens of micrometer level
should be investigated in more detail. Eventually, random misalignments of the beamline elements
should be replaced by a proper ground motion model
for the actual ILC site in the north of Japan. Many studies performed here depend on the beam energy.
In general it is expected that maintaining a permille-level polarisation measurement
becomes more difficult with increasing beam energy,
while at lower energies, e.g.\ at the Higgs or $t\bar{t}$ threshold, the impact of the collision effects
is reduced. This, however, needs to be quantified. Finally, in order to eliminate the uncertainty
from assuming a certain spin configuration at the starting point of the simulation,
a full start-to-end spin simulation of the ILC would be highly desirable.

\section*{Acknowledgement}
We thankfully acknowledge the support by the BMBF Verbundforschung ``Spin Optimierung'' and by the DFG via
the Emmy-Noether-Grant Li/1560-1. We thank Karsten B\"u\ss er, Mathias Vogt, 
Peter Sch\"uler, Anthony Hartin, Desmond Barber, Ken Moffeit, Mike Woods and Yuri Nosochkov for many helpful discussions over the last years.

\appendix

\section{Appendix}      \label{sec:appendix}
All results presented in this paper are based on \cite{thesisMoritz}. Some simulations have been rerun with the following modifications:
\begin{itemize}
   \item The initial spin configurations are slightly different (see section~7.1.2 in \cite{thesisMoritz} and section~\ref{sec:crosscal} in this paper). The discrepancy is however much smaller than the difference between the samples \samtdr\ and \samtdrstar\ in section~\ref{sec:lumipol}, since the angular divergence at the beginning of the lattice is only 1\,mrad.
   \item While the collisions are simulated with the TDR beam parameters in this paper (samples \samtdr\ and \samtdrstar), the corresponding simulations in \cite{thesisMoritz} are performed with RDR beam parameters and an increased bunch charge, such that the amount of beamstrahlung produced in the collisions is approximately the same as for collisions with TDR beam parameters (sample $\Upsilon_{TDR}$).
   \item For the collision effects, some data samples have been simulated without beamstrahlung and synchrotron radiation. In \cite{thesisMoritz}, the initial beam energy spreads are still present (sample no BS,SR), while in this paper also the initial beam energy spreads are set to zero (samples \samrdrfixedEnergy, \samtdrfixedEnergy\ and \samtdrstarfixedEnergy).
   \item Misalignments: In contrast to this paper, the corresponding simulations in \cite{thesisMoritz} include bunch rotation at the IP (crab cavities) and energy losses due to synchrotron radiation.
   \item Measurable Polarisation: In \cite{thesisMoritz}, the crossing angle of the laser beam has not been taken into account. Therefore, it is defined by a cut on the radius around the bunch center rather than by a cut on the horizontal distance.
\end{itemize}

\newcommand{\reftitle}[1]{``#1,''}

\end{document}